\documentclass[a4paper,11pt]{article}
\usepackage{jcappub} % for details on the use of the package, please see the JINST-author-manual
\usepackage{amsmath}
\usepackage{comment}
\usepackage{xcolor}
\usepackage{graphicx}
\title{UHECR doublets and their conditional association with nearby radio galaxies}
\author{V.~Barbosa~Martins}
\affiliation{Instituto de Física, Universidade Federal de Goiás, 74690-900, Goiânia, GO, Brazil}
\affiliation{Ruhr University Bochum, Faculty of Physics and Astronomy, Astronomical Institute (AIRUB), 44780 Bochum, Germany}

\emailAdd{barbosamartins@ufg.br}

\abstract{The origin of ultra-high-energy cosmic rays (UHECRs) remains a fundamental question in astroparticle physics. While localized $\sim$3$\sigma$ correlations with active galactic nuclei and starburst galaxies have been reported using time-integrated analyses, we propose and implement a spatiotemporal multiplet search method utilizing a pre-defined fixed window of $3^\circ$ and 15 days—a kinematic filter designed to isolate high-rigidity particles and keep chance coincidences low. Applying this method to 16 years of Pierre Auger Observatory data, we identify 28 UHECR multiplets (doublets) above 32~EeV. We backtrack these trajectories using three Galactic magnetic field models across eight distinct nuclear species. Testing the backtracked directions against the ten nearest bright radio galaxies yields an overall post-trial significance of $5.8\sigma$, conditioned on the individual best-fit associations. Specifically, we find a 4.5$\sigma$ conditional post-trial significance for the joint association of 8 of these multiplets with the Fornax~A region alone. These results point to radio galaxies, with a strong contribution from Fornax A, as long-term accelerators of heavy ($Z>3$) UHECRs—possibly within the mildly relativistic backflows of their extended radio lobes—detected at Earth primarily as independent secondary fragments above 32 EeV.}
\usepackage{xparse}
\usepackage{caption} % Allows \caption* for sub-labels if needed
\NewDocumentCommand{\MultipletPage}{m m m m}{
    \begin{figure}[p]
        \centering
        
        \begin{minipage}{0.49\textwidth}
            \centering
            \makebox[\linewidth][c]{\includegraphics[width=\linewidth]{figures/backtrack_M#1_markers.pdf}}
        \end{minipage}
        \hfill
        \if\relax\detokenize{#2}\relax\else
            \begin{minipage}{0.49\textwidth}
                \centering
                \makebox[\linewidth][c]{\includegraphics[width=\linewidth]{figures/backtrack_M#2_markers.pdf}}
            \end{minipage}
        \fi

        \vspace{0.5cm}

        \if\relax\detokenize{#3}\relax\else
            \begin{minipage}{0.49\textwidth}
                \centering
                \makebox[\linewidth][c]{\includegraphics[width=\linewidth]{figures/backtrack_M#3_markers.pdf}}
            \end{minipage}
        \fi
        \hfill
        \if\relax\detokenize{#4}\relax\else
            \begin{minipage}{0.49\textwidth}
                \centering
                \makebox[\linewidth][c]{\includegraphics[width=\linewidth]{figures/backtrack_M#4_markers.pdf}}
            \end{minipage}
        \fi
        
        \caption{Backtracking simulations for multiplet\if\relax\detokenize{#2}\relax\else s\fi\ M#1%
        \if\relax\detokenize{#2}\relax\else, M#2\fi%
        \if\relax\detokenize{#3}\relax\else, M#3\fi%
        \if\relax\detokenize{#4}\relax\else, and M#4\fi. 
        The layout and markers follow the same convention as Fig.~\ref{fig:backtracking_grid}. 
        The cyan star marks the closest source, not necessarily the associated one (check Table~\ref{tab:individual_results}).}
        \label{fig:app_multiplets_#1}
    \end{figure}
    \clearpage
}

\begin{document}
\maketitle
\flushbottom

\section{Introduction}
\label{sect:introduction}

Ultra-high-energy cosmic rays (UHECRs) are the most energetic particles
observed in nature, reaching energies well above $10^{18}$~eV and, in the most extreme events, exceeding $10^{20}$~eV~\cite{Kotera2011,Aloisio2017}. Despite more than a century having passed since the discovery of cosmic radiation, the astrophysical sources responsible for accelerating particles to such extreme energies remain one of the most profound open questions in modern astroparticle physics~\cite{AlvesBatista_2019}. The identification of these sources carries deep implications not only for high-energy astrophysics but also for our understanding of particle acceleration mechanisms, the large-scale structure of the universe, and the nature of cosmic magnetic fields.

The challenge of identifying UHECR sources is fundamentally tied to the
deflections these charged particles undergo as they traverse the Galactic Magnetic Field (GMF) and the Extragalactic Magnetic Field (EGMF). Unlike photons and neutrinos, which travel in straight lines, cosmic rays carry electric charge and are continuously deflected along their propagation paths, with the deflection scaling inversely with rigidity $R = E/Z$, where $E$ is the energy and $Z$ is the charge number of the nucleus. For particles of sufficiently high rigidity, the deflection by the GMF can amount to several tens of degrees~\cite{JF12,PT11,TF17}, rendering a direct angular correlation between detected events and their putative sources practically impossible without a careful accounting of the intervening magnetic environment.

Progress toward solving the source problem has nonetheless been made on
several observational fronts. The Pierre Auger Observatory in Argentina, the largest ground-based cosmic ray detector ever built, has reported compelling evidence for a large-scale dipolar anisotropy in the arrival directions of events above 8~EeV, pointing away from the Galactic center and consistent with an extragalactic origin~\cite{Aab2017dipole}. Furthermore, studies of localized correlations with catalogued Active Galactic Nuclei (AGN), starburst galaxies, and jetted AGN have shown directional excesses at the level of several standard deviations, though none of these analyses has yet reached the threshold of a definitive source identification~\cite[e.g., ][]{AugerAGN2018,Auger2022}.

Radio galaxies, a subclass of AGN characterized by powerful relativistic jets that extend far beyond the optical confines of the host galaxy, are among the most widely favored candidate sources of UHECRs~\cite{Hardcastle2020}. The relativistic jets and giant radio lobes of these objects can sustain magnetic fields and velocity shocks over enormous spatial scales, conditions that are well suited for particle acceleration through diffusive shock acceleration and related mechanisms~\cite{Matthews2019}. Centaurus~A, the nearest radio galaxy at a distance of approximately 4.1~Mpc, has long been regarded as a particularly promising source candidate, and a directional excess toward its position has been reported in multiple Pierre Auger Collaboration's analyses~\cite{AugerAGN2018}. Fornax~A and other members of the local radio galaxy population, compiled in the all-sky catalog \cite{vanVelzen_2012}, likewise represent compelling candidates given their proximity and radio luminosity. Recent results confirm that the source horizon for UHECRs lies within 20--100\,Mpc, a distance range dependent on the mass composition \cite{Lang}.

A further layer of complexity in source identification arises from the nuclear composition of arriving UHECRs. Measurements of the depth of shower maximum ($X_\mathrm{max}$) at the
Pierre Auger Observatory and the Telescope Array suggest a gradual transition toward a heavier average composition at the highest energies~\cite{2024JCAP...01..022A, 2025PhRvL.134b1001A}. This compositional picture has profound consequences for any directional analysis, because a heavy nucleus with atomic number $Z$ and energy $E$ undergoes a magnetic deflection $Z$ times larger than a proton at the same energy, substantially broadening the apparent arrival direction distribution around any true source. Moreover, heavy nuclei are subject to photo-disintegration interactions with the cosmic microwave background and the extragalactic background
light~\cite{2023APh...15202866K}, which progressively strip nucleons from the parent nucleus during propagation \cite{Khan2005}, producing a cascade of lighter secondary nuclei that arrive at Earth with a slightly modified direction still approximately tracing the original trajectory.

Because magnetic deflections scale strictly with rigidity, particles possessing the exact same rigidity follow very similar trajectories from their source to Earth. Conversely, particles with different rigidities will experience diverging path lengths, leading to macroscopic differences in their arrival times. Consequently, the observation of spatio-temporal multiplets—independent mutually exclusive events arriving from a tightly clustered spatial region within a narrow temporal window—requires a strict matching of rigidities among the detected particles, considering they originate from the same source. If the rigidities differed even slightly, the intervening magnetic fields would act as a spectrometer, separating their arrival times by years or millennia, through different paths. This rigid matching condition isolates specific astrophysical scenarios: the particles comprising a multiplet may be independent primaries emitted by the same source with the same rigidity, or they may be secondary fragments produced during intergalactic propagation that happened to reach similar rigidities.

The spatio-temporal search departs from conventional time-integrated approaches. While previous multiplet searches by the Pierre Auger Collaboration \cite{Auger_multiplets} integrated data over several years to look for a correlation between arrival direction and the inverse of energy, such long observation windows inherently increase the probability of chance alignments from largely deflected particles. By imposing strict angular and temporal constraints, our method utilizes magnetic rigidity to filter out unassociated cosmic rays. This dual constraint suppresses chance coincidences when identifying doublets, increasing the likelihood of an astrophysical correlation between them.

In this work, we apply the doublet search methodology to 16 years of publicly available UHECR data from the Pierre Auger Observatory~\cite{dataset_auger}. Doublets are identified using a pre-defined spatial window of $3^\circ$ and a temporal window of 15 days. The backtracking of doublets through the GMF is performed with the CRPropa 3 simulation framework~\cite{crpropa3}, employing the JF12~\cite{JF12}, PT11~\cite{PT11}, and TF17~\cite{TF17} GMF models to bracket systematic uncertainties, and scanning over eight representative nuclear species from protons to iron. The association between backtracked doublet positions and the ten brightest local radio galaxies within 100~Mpc is assessed through a stacking analysis, with statistical significance evaluated against a null hypothesis that accounts for the anisotropic matter density of the local large-scale structure as traced by the 2MASS Redshift Survey~\cite{Huchra2012}.

This paper is organized as follows: Section~\ref{sec:methods} describes the dataset, the doublet identification algorithm, the backtracking methodology, the radio galaxy catalog, and the full statistical framework; Section~\ref{sec:results} presents the results of the multiplet identification and source association analysis; Section~\ref{sect:discussions} discusses the physical interpretation of the findings; and Section~\ref{sect:conclusions} summarizes the conclusions.

\section{Methods}
\label{sec:methods}

\subsection{The dataset}
We utilize in our analysis the 2635 events above 32 EeV from 2004 to 2020 as provided by the Pierre Auger Collaboration \cite{dataset_auger}. The available information are reconstructed energy, arrival direction, and integrated exposure. We highlight that no information on the shower parameters, such as the maximum depth were available to estimate the nuclear species of the cosmic rays. In addition to that, we use the instrument's exposure map as a proxy to assess the instrument's sensitivity as function of arrival direction, and the exposure time to assess the rate of data acquisition over time.

\subsection{Finding and assessing doublets}
\label{sec:find_multiplets}
Our methodology is centered on the identification of multiplet (specifically doublet) events, defined as pairs of UHECRs arriving within a shared spatio-temporal window. We define a priori and adopt a spatial radius of $3^\circ$ and a temporal window of 15 days, driven by both instrumental and physical constraints. Computationally, we extract these multiplets by performing an efficient spatial nearest-neighbor search using a $k$-d tree algorithm, followed by a temporal filter and graph-based connected component analysis to group the associated events, implemented via the SciPy library~\cite{scipy2020}. The $3^\circ$ angular separation safely encapsulates the event reconstruction resolution of the Pierre Auger Observatory at these energies ($\lesssim 1^\circ$)~\cite{dataset_auger}. The 15-day temporal window serves primarily to suppress chance coincidences. By limiting the allowed arrival time difference, we minimize the probability of falsely pairing uncorrelated events, thereby increasing the chances that the identified multiplets share the same astrophysical origin. This spatiotemporal window works as a kinematic filter. Because turbulent magnetic scattering disperses low-rigidity events, enforcing simultaneous spatial and temporal constraints filters for high-rigidity particles capable of preserving a certain degree of directional coherence. Due to the stochastic nature of magnetic deflections, this selection cannot guarantee the purity of individual multiplets; highly deflected particles will inevitably satisfy the matching criteria by chance. The expected rate of these chance coincidences is explicitly quantified below. The stability of the results against these parameter choices is detailed in Appendix~\ref{app:doublets_systematics}.

To quantify the probability of identifying doublets purely by chance, we conduct two distinct statistical analyses: a global coincidence test and a local coincidence test. The global test assesses the significance of finding the total number of doublets identified in the data by chance. We generate a million Monte Carlo realizations of the entire UHECR dataset, distributing events according to the instrument's spatial and temporal exposure. In each realization, we search for doublets using the same spatiotemporal criteria applied to the real data. By building a distribution of the number of accidental doublets expected from the Monte Carlo simulation, we can compare it to our observed count; a significantly lower mean in the simulated distribution compared to the observation indicates that the doublets in the dataset are likely physical rather than chance coincidences. In this purely statistical check, we do not yet consider any information on the potential sources or on the magnetic field deflections, which is done in the following section.

The second test is designed to estimate the probability that a second event would serendipitously fall within the defined angular and temporal window of the first event, for each doublet. Crucially, this evaluation imposes an energy-ordering condition: a coincidence is only counted if the energies of the simulated pair, sampled from the dataset, are at least as high as those of the observed doublet (requiring $E_{\mathrm{sim}}^{\mathrm{min}} \ge E_{\mathrm{obs}}^{\mathrm{min}}$ and $E_{\mathrm{sim}}^{\mathrm{max}} \ge E_{\mathrm{obs}}^{\mathrm{max}}$). To achieve high precision, we simulate $10^8$ independent events, accounting for spatial sensitivity via the instrument's exposure map and temporal sensitivity by using the differential exposure, derived from the integrated exposure provided for each event in the dataset. We then count the frequency with which a simulated event satisfies the selection criteria relative to the fixed event. This methodology accounts for operational fluctuations between the reported events, such as maintenance shutdowns or upgrades. Although this approach assumes the spatial profile of the exposure map is constant over time, the assumption holds well because array expansions and upgrades at the Pierre Auger Observatory are not expected to significantly alter the overall spatial acceptance.

\subsection{Backtracking of doublets through the Galactic magnetic field}
\label{sec:backtracking}
Once doublets are identified in the dataset, we investigate their origin by accounting for the deflection of UHECRs by the GMF. We neglect as a first approximation the EGMF, as it is assumed to be sub-dominant over the relatively short distances to nearby sources \cite{AlvesBatista_2019}, although we study  the influence of this assumption in Appendix~\ref{app:egal}. Furthermore, deflections at the source are not analyzed here, as our primary focus is backtracking the arrival directions observed at Earth rather than estimating the intrinsic UHECR emissivity at the source.

To reconstruct the source position, we employ the backtracking technique using \textit{CRPropa} 3 \cite{crpropa3}. This method relies on the physical symmetry that the trajectory of an anti-nucleon propagating backwards from Earth through a static magnetic field is identical to the forward trajectory of a nucleon. The simulation tracks particles up to a Galactocentric distance of 20\,kpc. This boundary is chosen to ensure the inclusion of the full extent of the Galactic halo's coherent magnetic field structures, beyond which the field strength decays to negligible levels.

To bracket the systematic uncertainties associated with the GMF geometry, we use three distinct field models available in \textit{CRPropa}: Jansson and Farrar \cite[JF12, ][]{JF12}, Pshirkov and Tinyakov \cite[PT11, ][]{PT11}, and Terral and Ferrière \cite[TF17, ][]{TF17}. These models offer varying representations of the large-scale regular field, ranging from different spiral arm geometries to X-shaped halo structures.

Crucially, UHECRs undergo stochastic scattering due to small-scale magnetic irregularities. To strictly isolate the impact of the large-scale field geometry, we employ a unified turbulence background across all models. We utilize the random field component of the JF12 model, which is generated via a seeded randomization process (randomTurbulent) scaling with the local relativistic gas density. For the JF12 regular field scenarios, we include both the isotropic turbulent component and the anisotropic striated random field (randomStriated). However, when testing the PT11 and TF17 geometries, we explicitly disable the striated component and superimpose only the small-scale turbulent field onto the respective regular fields. This ensures that the scattering power remains consistent across all tested geometries while avoiding geometric conflicts with the spiral arm definitions of the PT11 and TF17 models.

For every doublet, we conduct 100 independent backtracking simulations per field model by injecting antiparticles with charge $-Ze$ and velocity vectors inverted relative to the observed UHECR arrival directions. The trajectories are propagated using the Cash-Karp numerical integration scheme with an adaptive step size ranging from 0.1 pc to 50 pc. The upper limit is deliberately chosen to remain on the order of the typical coherence length of the turbulent Galactic magnetic field, while the lower limit ensures high precision in regions with strong local gradients. The simulation for a given particle terminates when it crosses the Galactic Halo boundary, defined as a geocentric sphere of 20 kpc, beyond which the Galactic field strength becomes negligible. A maximum trajectory length of 500 kpc is imposed as a computational fail-safe to prevent infinite loops for particles undergoing severe magnetic trapping. We perform this analysis for a representative set of nuclei—protons ($^{1}$H), helium ($^{4}$He), lithium ($^{7}$Li), beryllium ($^{9}$Be), boron ($^{11}$B), nitrogen ($^{14}$N), oxygen ($^{16}$O), and iron ($^{56}$Fe)—spanning the compositional mass groups typically inferred at these energies. Given the high magnetic rigidity of the events in our sample, their propagation is predominantly ballistic rather than diffusive. Consequently, the turbulent dispersion envelope is relatively narrow, and 100 stochastic realizations provide a statistically sufficient representation of the source variance, especially for lighter nuclei where the method is most efficient. We confirmed this empirical stability by observing no significant deviation in the spatial spread when incrementally increasing the number of realizations from 20 to 50 and then to 100.

\subsection{Catalog of radio galaxies}
To identify potential astrophysical counterparts for the observed UHECR doublets, we focus on radio galaxies, a class of Active Galactic Nuclei (AGN) featuring powerful relativistic jets widely postulated as acceleration sites for ultra-high energy cosmic rays. We employ the catalog of local radio galaxies compiled by van Velzen et al. (2012) \cite{vanVelzen_2012}. From this catalog, we select the ten brightest sources within a distance horizon of $D < 100$\,Mpc, prioritizing those with the highest radio flux. The 100\,Mpc cut is physically motivated by the GZK horizon for particles above our 32\,EeV threshold, beyond which the flux is severely attenuated by energy losses. Furthermore, selecting the top ten sources by radio flux provides a robust, predefined proxy for the most luminous local accelerators. The selected candidates and their fluxes are listed in Table~\ref{tab:radio_galaxies}.

\begin{table}[htbp]
\centering
\begin{tabular}{l|cccc}
\hline
\textbf{Name} & \textbf{Flux (Jy)} & \textbf{Dist (Mpc)} & \textbf{z}\\
\hline
Cen A        & 1803.15 & 8.1  & 0.0018 \\
Fornax A     & 169.00  & 26.1 & 0.0059 \\
M 87         & 147.45  & 19.4 & 0.0044 \\
IC 4296      & 26.45   & 55.8 & 0.0125 \\
NGC 1275     & 22.83   & 78.8 & 0.0176 \\
NGC 5090     & 12.83   & 51.0 & 0.0114 \\
UGC 7360     & 10.44   & 33.3 & 0.0075 \\
2MASX J1254  & 7.71    & 69.1 & 0.0154 \\
CGCG 114-025 & 7.33    & 75.8 & 0.0169 \\
NGC 0315     & 6.63    & 74.0 & 0.0165 \\
\hline
\end{tabular}
\caption{The ten brightest local radio galaxies ($D < 100$ Mpc) selected from the catalog of \cite{vanVelzen_2012}. Note: The cosmological distance is calculated from the redshift using the Planck18 cosmology model in \textit{astropy.cosmology} \cite{astropy}.\label{tab:radio_galaxies}}
\end{table}

It is worth noting the spatial configuration of the two closest sources in our sample, Cen A and NGC 5090. These galaxies are separated by only $\sim 1^{\circ}$ on the sky, which is approximately the systematic uncertainty on the direction reconstruction of the instrument \cite{Bonifazi:2009ar}, despite their distinct distances ($D \approx 8.1$\,Mpc and $D \approx 51.0$\,Mpc, respectively). While Cen A dominates the local radio flux ($S_{\nu} \approx 1800$\,Jy at 1.4 GHz), NGC 5090 is also a strong radio emitter ($S_{\nu} \approx 13$\,Jy) with substantially larger radio lobes.

\subsection{Statistical Significance Assessment}
\label{sec:significance}
To quantify the correlation between the backtracked UHECR multiplets and the candidate radio galaxies, we first establish a probabilistic relation for each individual multiplet-nucleus-source combination. As detailed in the previous section, for every multiplet, every magnetic field model, and every tested nuclear species $Z$, we perform 100 independent backtracking simulations.

\subsubsection{Probability Maps and Optimized Scoring}
For each multiplet $i$ and assumed nuclear species $Z \in \mathcal{Z} = \{ \text{p, He, Li, Be, B, N, O, Fe} \}$, we represent the backtracked arrival directions as a continuous probability density function (PDF), $P_{i,Z}(\hat{n})$. This PDF is constructed by fitting a 2D spherical Gaussian to the coordinates derived from 100 backtracking simulations. This approach effectively characterizes the positional spread induced by the turbulent component of the GMF. This is particularly robust for lighter nuclei, which typically exhibit smaller angular dispersions and probe more localized regions of the GMF. If the PDFs for a multiplet do not overlap with any of the catalog sources, no source is associated with it.

We treat the nuclear number and the specific source association as free parameters and define the association score $S_{i}$ for a multiplet $i$ by maximizing the likelihood across all candidate sources $j$ in the catalog $\mathcal{C}$ and all tested isotopes $Z$:

\begin{equation}
    S_{i} = \max_{j \in \mathcal{C}} \left( \max_{Z \in \mathcal{Z}} \left[ P_{i,Z}(\hat{n}_j) \right] \right)
\end{equation}

The specific source-nucleus pair $(j, Z)$ that maximizes this probability is adopted as the most likely physical association for the multiplet. This procedure effectively selects the candidate source and the physically allowed trajectory that collectively minimize the angular separation between the backtracked events and the catalog.

\subsubsection{Null Hypothesis and Significance Testing}
To assess the statistical significance of the observed associations, we employ a Monte Carlo approach. Rather than generating and numerically backtracking massive ensembles of random multiplets—a computationally expensive task—we keep the real backtracked coordinates fixed and instead randomize the source positions. We generate $10^10$ synthetic catalogs of 10 sources each, drawing their coordinates from the 2MASS Redshift Survey (2MRS) catalog \cite{Huchra2012} to ensure the null hypothesis preserves the realistic clustering of local matter.

We utilize the same UHECR probability maps derived from the backtracked events to evaluate the association score for every synthetic source-multiplet-nucleus combination. Analogous to the treatment of the observed dataset, we maximize this score over the set of tested isotopes, thereby associating a specific best-fit nuclear species $Z$ with each random source in the synthetic catalog. This method effectively internalizes the trial correction for the number of nuclei tested; a high score in the data is deemed significant only if it exceeds the maximum likelihoods obtainable by random sources optimizing their fit to any of the considered nuclear distributions.

The pre-trial probability ($p_{pre}$) is defined as the fraction of synthetic realizations in which the maximized association score exceeds the score observed in the actual dataset. We then apply a penalization exclusively for the spatial look-elsewhere effect arising from the scan of the $N_{src}=10$ candidate sources. The post-trial significance ($p_{post}$) is derived via the Šidák correction:

\begin{equation}
    p_{post} = 1 - (1 - p_{pre})^{N_{src}}
    \label{eq:spatial_correction}
\end{equation}
where $N_{src}=10$.

To assess the physical plausibility of the associations, we calculated the survival probability ($f_{\text{surv}}$) of the best nucleus in each multiplet. We performed 1D Monte Carlo simulations using CRPropa 3. The simulations accounted for the full range of relevant interactions: photodisintegration, photo-pion production, electron pair production, and nuclear decay. We modeled the extragalactic background light (EBL) using the Gilmore 2012 model \cite{Gilmore:2012} and employed TALYS cross-sections \cite{Koning} for nuclear interactions. For each particle in a significant doublet, we injected 1,000 realizations at the source distance. Crucially, we initialized the simulation at the source using the energy observed at Earth. Since UHECRs lose energy during propagation (meaning the true injection energy at the source must have been higher), and since interaction horizons generally shrink with increasing energy, this approach minimizes the interaction probability. Consequently, the resulting survival fractions derived here should be interpreted as upper limits on the true survival probability.

\subsection{Source Stacking and Catalog Analysis}
\label{sec:stacking}
While the previous sections assessed the significance of individual nucleus-source associations, the primary strength of this analysis lies in the ``stacking'' of signals. Since the detection of each UHECR multiplet constitutes an independent event, we can combine the statistical weight of multiple multiplets associated with the same candidate source to test for a collective correlation. The pertinent question is whether the specific physical model suggested by the data—specifically, the unique combination of multiplet-nucleus associations for each source—is statistically distinguishable from chance.

\subsubsection{Single-Source Stacking}
For a specific radio galaxy $j$, let $\mathcal{M}_j$ be the set of multiplets associated with it (i.e., those for which source $j$ provided the highest score in the assignment step). We define the stacked score $\mathcal{T}_j$ as the sum of the optimized association scores for each multiplet $i$ in this set:
\begin{equation}
    \mathcal{T}_j = \sum_{i \in \mathcal{M}_j} \left( \max_{Z} S_{i,j,Z} \right)
\end{equation}

We assess the significance of the stacked multiplets for each source by comparing the observed sum $\mathcal{T}_j$ against the scores derived from synthetic source catalogs. Crucially, this comparison employs a conditional hypothesis: for the synthetic trials, the nuclear species for each multiplet is fixed to the specific $Z_{\text{best}}$ identified in the individual analysis. The pre-trial probability $p_{\text{pre}}$ is defined as the fraction of synthetic trials where the cumulative score under this fixed hypothesis equals or exceeds the observed $\mathcal{T}_j$. To strictly account for the look-elsewhere effect inherent in the model selection (scanning 10 sources and 8 nuclear species), we apply the Šidák correction (Eq.~\ref{eq:spatial_correction}) once with $N_\mathrm{iso}=8$ and once with $N_\mathrm{src}=10$.

\subsubsection{Whole-Catalog Stacking}
Finally, to test the hypothesis that the \textit{entire} selected catalog of radio galaxies acts as the dominant source population for UHECR doublets, we calculate a catalog test statistic $\mathcal{T}_{cat}$. This is defined as the grand sum of the stacking scores for all $N_{src}$ sources in the catalog:
\begin{equation}
    \mathcal{T}_{cat} = \sum_{j=1}^{N_{src}} \mathcal{T}_j
\end{equation}
We compare this observed grand total against the distribution of grand totals derived from the $10^{9}$ synthetic realizations, calculating, as before, the fraction of synthetic trials where $\mathcal{T}_{cat}$ is equal or exceeds the observed one. A high significance in this metric indicates that the set of local radio galaxies, as a population, correlates with the arrival directions of UHECR doublets significantly better than a random population of sources drawn from the local anisotropic matter density. Unlike the single-source case, this metric tests the single hypothesis of the "whole catalog," and thus requires no spatial trial correction. 

To ensure a conservative estimate, we further penalize this final significance by the probability that the multiplets themselves are chance coincidences (Sect.\ref{sec:find_multiplets}). To emphasize the correction applied to the spatial associations, we introduce a penalization factor based on the chance probability of the finding multiplets. We define a weighted probability as:

\begin{equation}P_{\text{final}} = P_{\text{spatial}}^{(1-P_{\text{chance}})}\end{equation}

\noindent In this expression, $P_{\text{spatial}}$ represents the raw probability of the spatial association between the source catalog and the identified multiplets from the preceding analysis step. The term $P_{\text{chance}}$ denotes the probability that the detected multiplets arise from random temporal coincidences rather than physical associations.This formulation explicitly uses the term $(1 - P_{\text{chance}})$ as a reliability weight. When a multiplet has a high probability of being a chance alignment (high $P_{\text{chance}}$), the exponent decreases, effectively penalizing and reducing the statistical significance of the spatial association. Conversely, for multiplets with high temporal reliability, the weight approaches unity, preserving the original spatial probability.

\section{Results}
\label{sec:results}

\begin{figure}[htbp]
    \centering
    \includegraphics[width=0.95\linewidth]{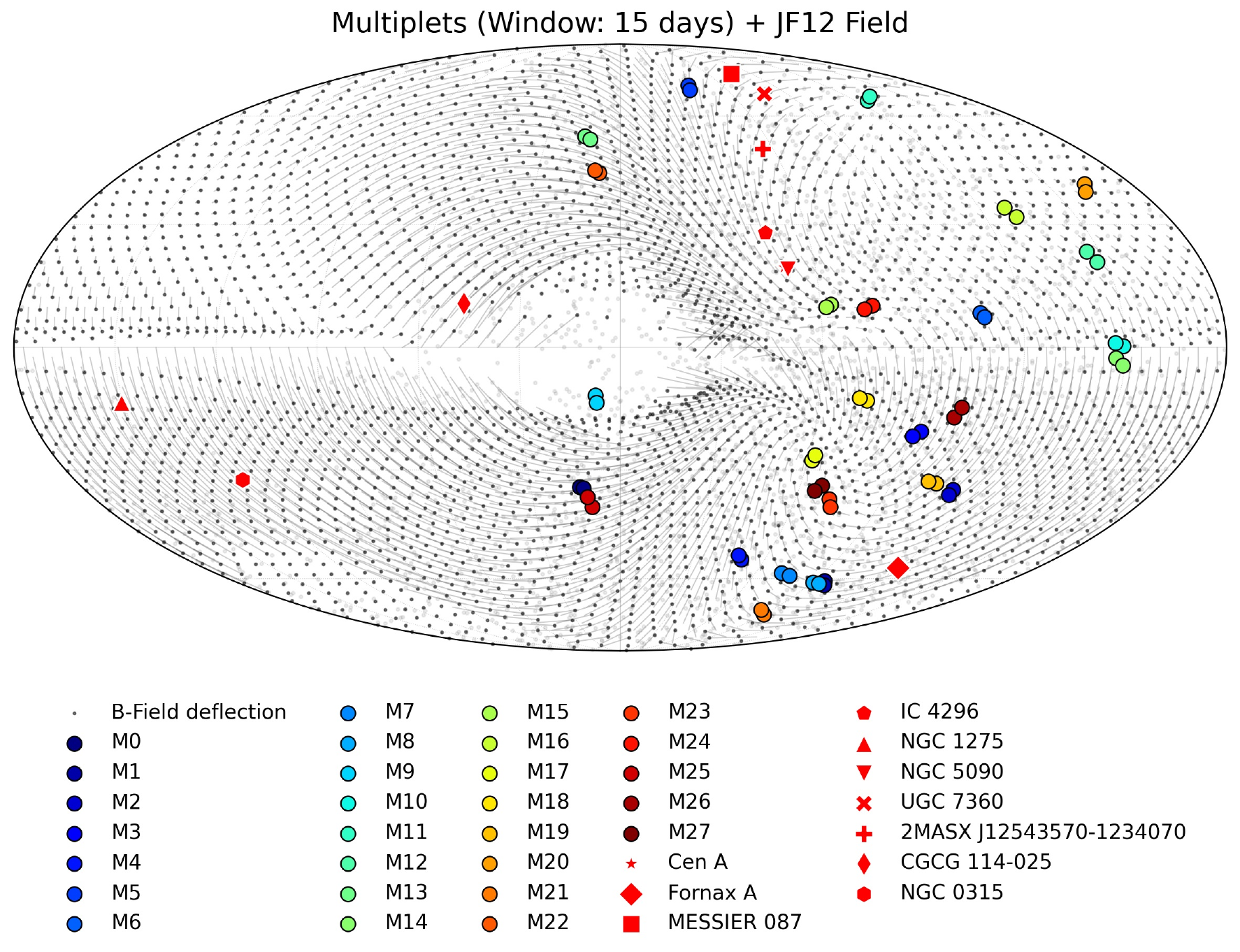}
    \caption{Sky map of the UHECR events in Galactic coordinates. 
    Events belonging to the identified spatiotemporal multiplets are shown as large colored circles. The remaining events from the dataset are plotted as small gray dots. 
    The positions of the ten selected radio galaxies are indicated by red markers. The faint black vectors illustrate the expected deflection direction for a $49$\,EeV proton in the JF12 GMF model.
    \label{fig:multiplet_map}}
\end{figure}

Applying the algorithm described in Section \ref{sec:find_multiplets} to the UHECR dataset, we identified a total of 28 doublets (and no higher-order multiplets). The events in the doublets are mutually exclusive, i.e., they do not repeat across doublets and occur at strictly unique time stamps. The spatial distribution of these event clusters is shown in Fig.~\ref{fig:multiplet_map}.

\begin{figure}[htbp]
    \centering
    % Left Panel: Global Significance (Histogram of Total Counts)
    \begin{minipage}{0.48\textwidth}
        \centering
        \includegraphics[width=\linewidth]{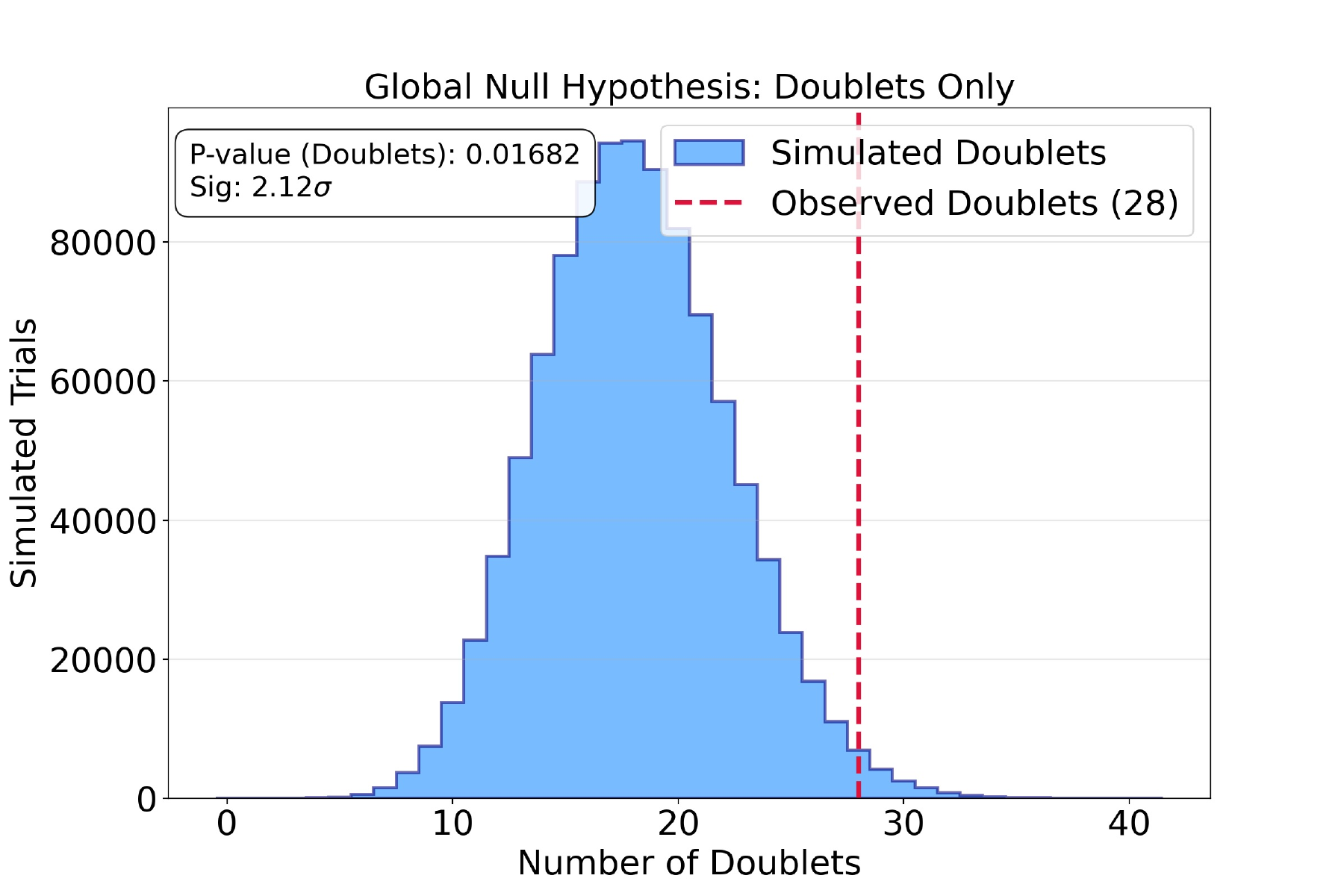}
        % \caption{Global Significance} % Optional sub-caption
    \end{minipage}
    \hfill
    % Right Panel: Local Significance (Distribution of Individual P-values)
    \begin{minipage}{0.48\textwidth}
        \centering
        \includegraphics[width=\linewidth]{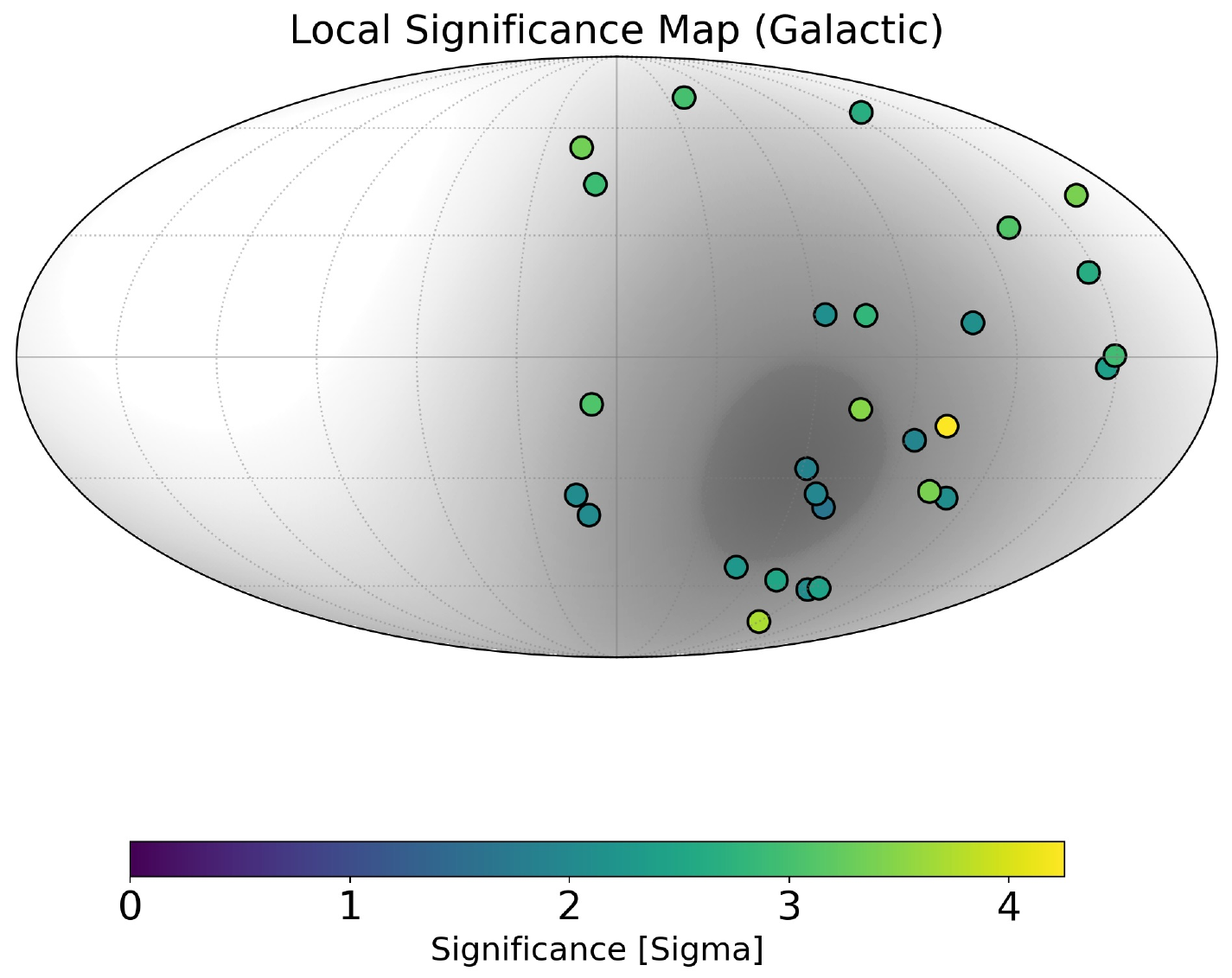}
        % \caption{Local Significance} % Optional sub-caption
    \end{minipage}
    
    \caption{Left: Distribution of the total number of doublets found in $10^6$ isotropic Monte Carlo realizations. The vertical red line marks the observed count ($N=28$) in the real dataset, corresponding to a significance of $2.1\sigma$ ($p \approx 0.017$). Right: Distribution of the local significances for the individual identified doublets. The gray map shows the instrument's exposure for reference.
    \label{fig:multiplet_sig}}
\end{figure}

The statistical significance of the observed clustering is evaluated through two complementary null-hypothesis tests. The probability of randomly observing 28 or more doublets is $p = 0.017$, corresponding to a significance of $2.1\sigma$, as illustrated in the distribution of simulated source realizations in Fig.~\ref{fig:multiplet_sig} (left).

We evaluate the local significance of each individual detection. This test calculates the chance probability of a second event falling within the spatiotemporal window ($\Delta t \leq 15$\,days, $\Delta \Psi \leq 3^{\circ}$) of a first event, accounting for the exposure at the specific coordinates and time of detection. As shown in Fig.~\ref{fig:multiplet_sig} (right), individual significances range from $1.6\sigma$ to 4.2$\sigma$. Furthermore, the multiplets labeled M1 and M8 (see Fig.~\ref{fig:multiplet_map}) coincide spatially within $3^{\circ}$ despite occurring at different times; the joint probability of this recurrence is $p = 1.4 \times 10^{-3}$, equivalent to a local significance of $3.0\sigma$ relative to the expectation under the null hypothesis.

Although the global excess ($2.1\sigma$) is moderate and dependent on the selected spatial and temporal windows (see e.g., Appendix~\ref{app:doublets_systematics}), the low probability of chance coincidence for the arrival of the second event in several individual doublets—coupled with the tight spatial coincidence of the M1 and M8 doublets—strongly indicates that a subset of these multiplets represents physically associated events originating from the same astrophysical source.

We proceed now to correct the doublet positions for deflections caused by the GMF. We applied the backtracking procedure detailed in Section \ref{sec:backtracking}, propagating antiparticles corresponding to the observed event energies backward through three distinct GMF models to account for model uncertainties.

While the backtracked regions vary depending on the specific GMF realization and rigidity assumptions, a directional alignment with the Fornax A is observed for 8 of the 28 identified multiplets. We show the backtracked trajectory for four of the doublets in Fig.~\ref{fig:backtracking_grid}. The remaining multiplets with association are shown in Appendix~\ref{app:all_multiplets}.

\begin{figure}[htbp]
    \centering
    % --- Row 1 ---
    \begin{minipage}{0.49\textwidth}
        \centering
        % \makebox[width of box][alignment]{content}
        % We set the box to \linewidth (size of minipage)
        % We set image to 1.15\linewidth (wider than minipage)
        \makebox[\linewidth][c]{\includegraphics[width=\linewidth]{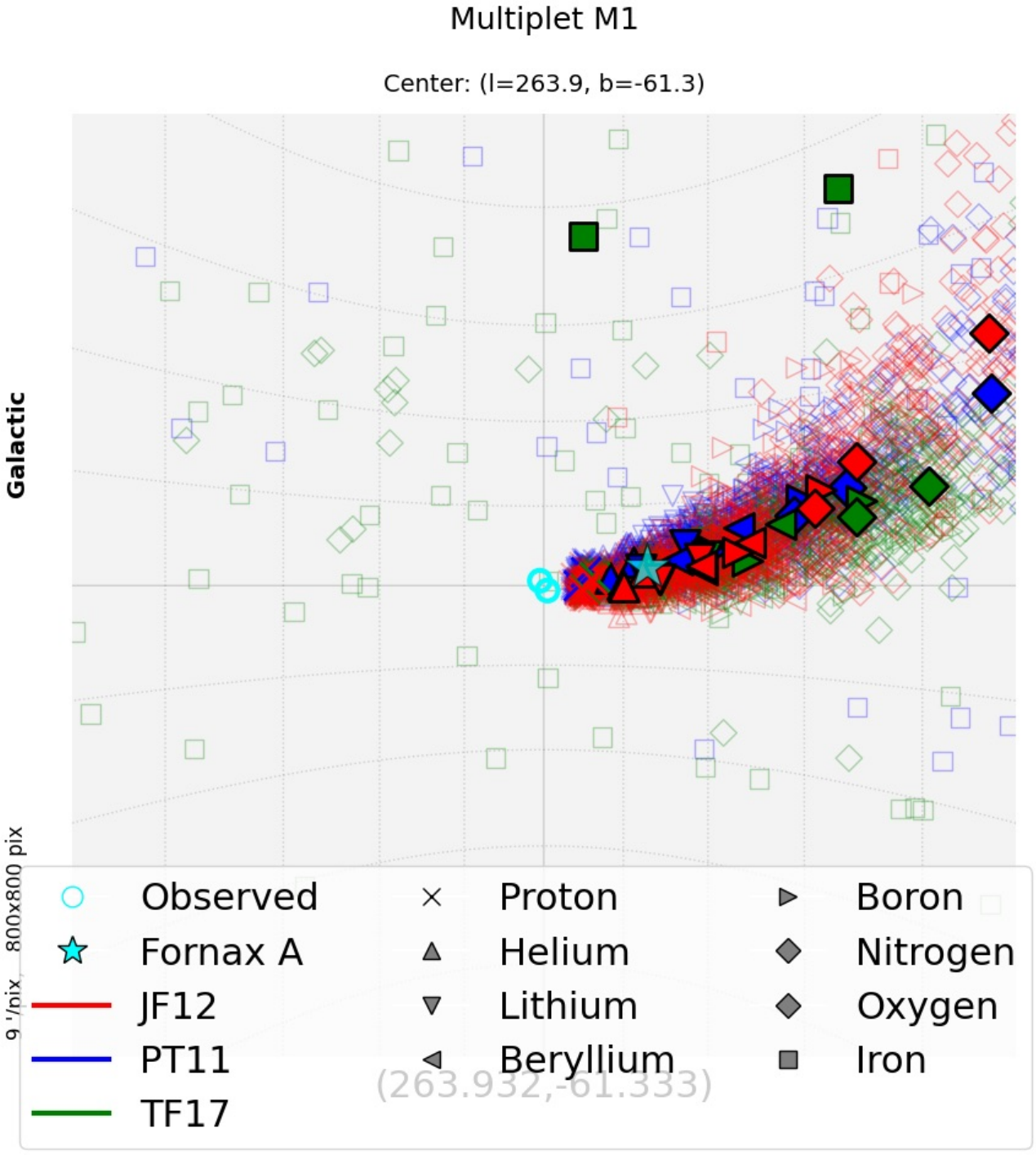}}
    \end{minipage}
    \hfill
    \begin{minipage}{0.49\textwidth}
        \centering
        \makebox[\linewidth][c]{\includegraphics[width=\linewidth]{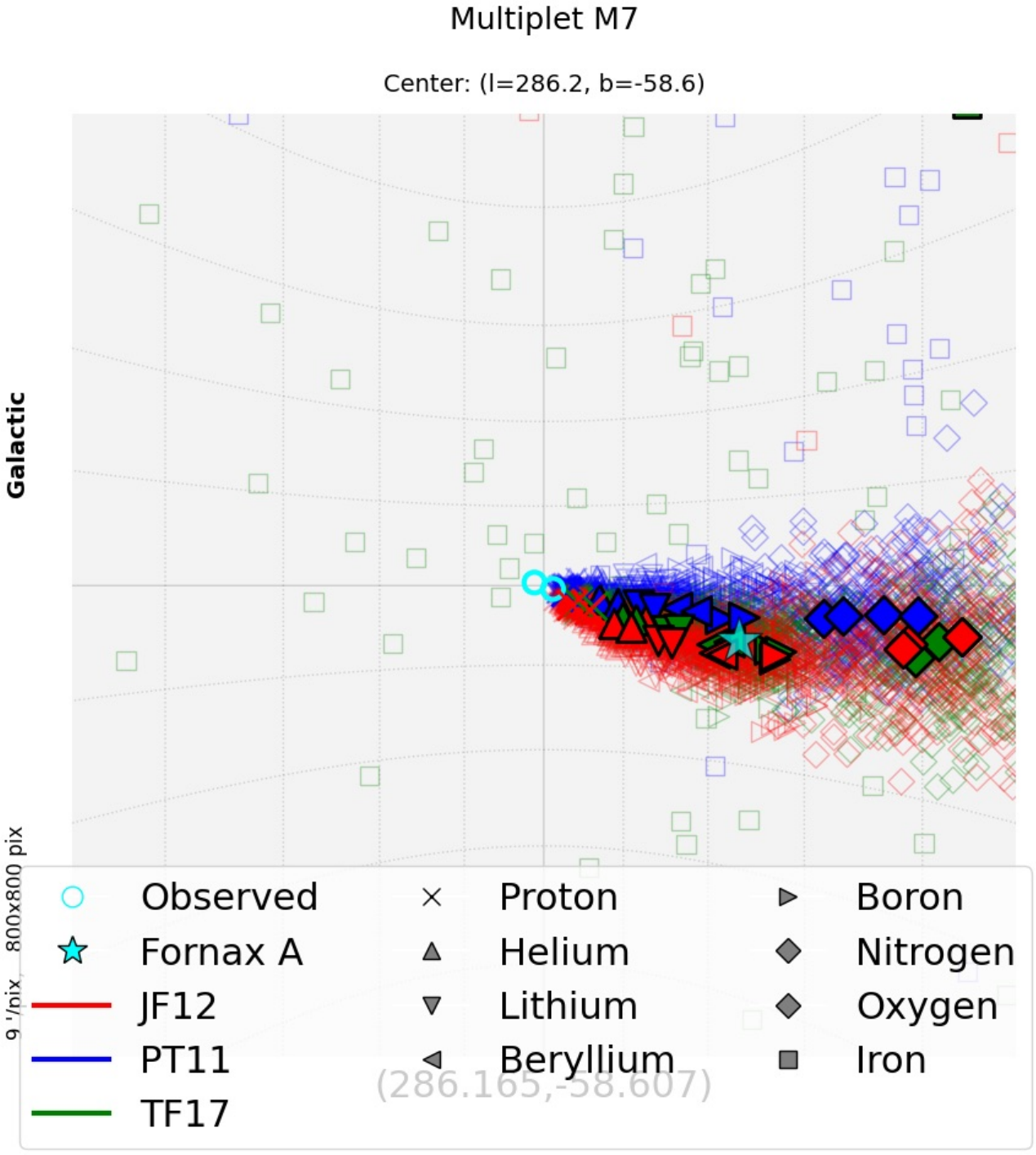}}
    \end{minipage}
    
    \vspace{0.1cm} % Adjust vertical space
    
    % --- Row 2 ---
    \begin{minipage}{0.49\textwidth}
        \centering
        \makebox[\linewidth][c]{\includegraphics[width=\linewidth]{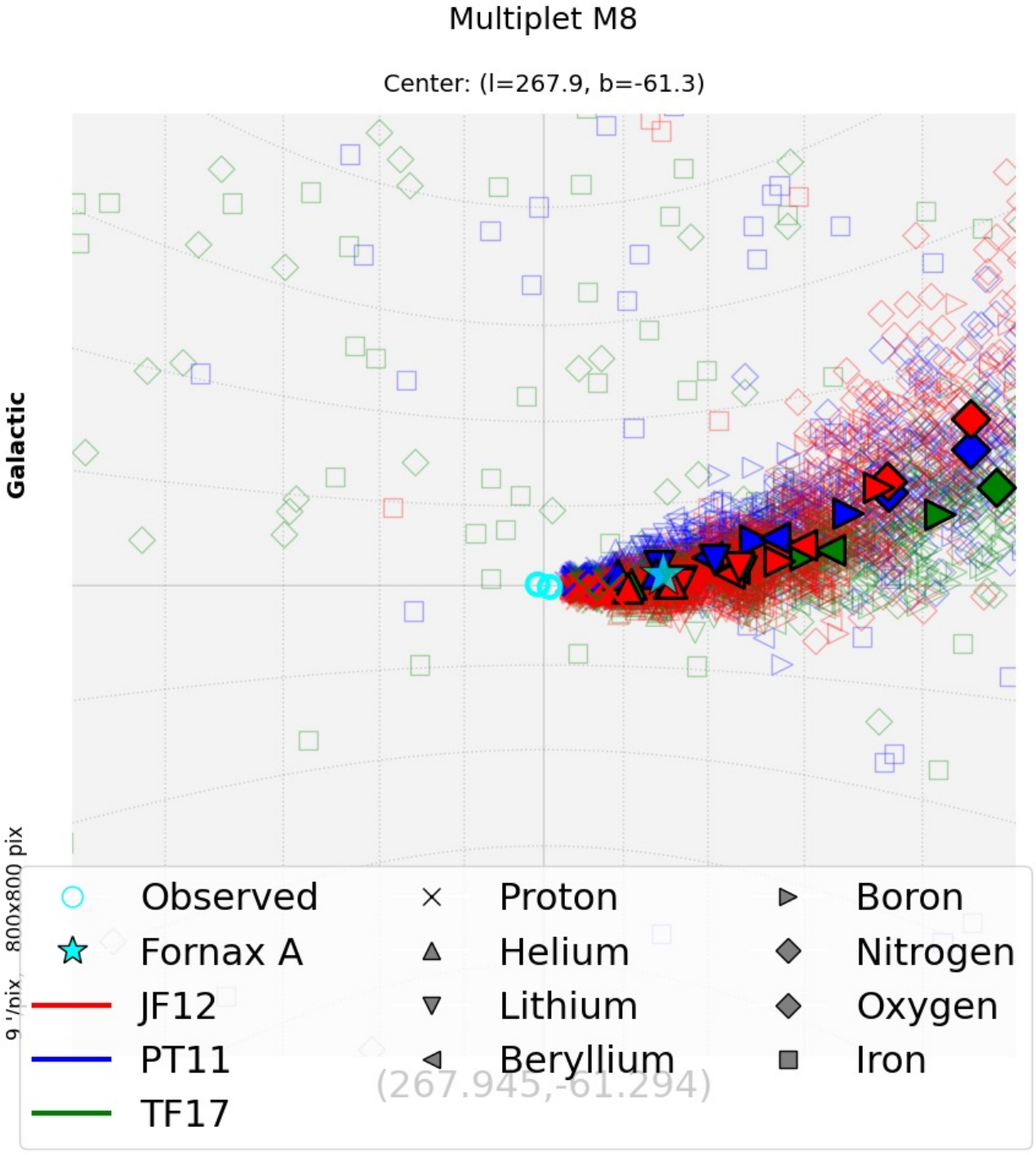}}
    \end{minipage}
    \hfill
    \begin{minipage}{0.49\textwidth}
        \centering
        \makebox[\linewidth][c]{\includegraphics[width=\linewidth]{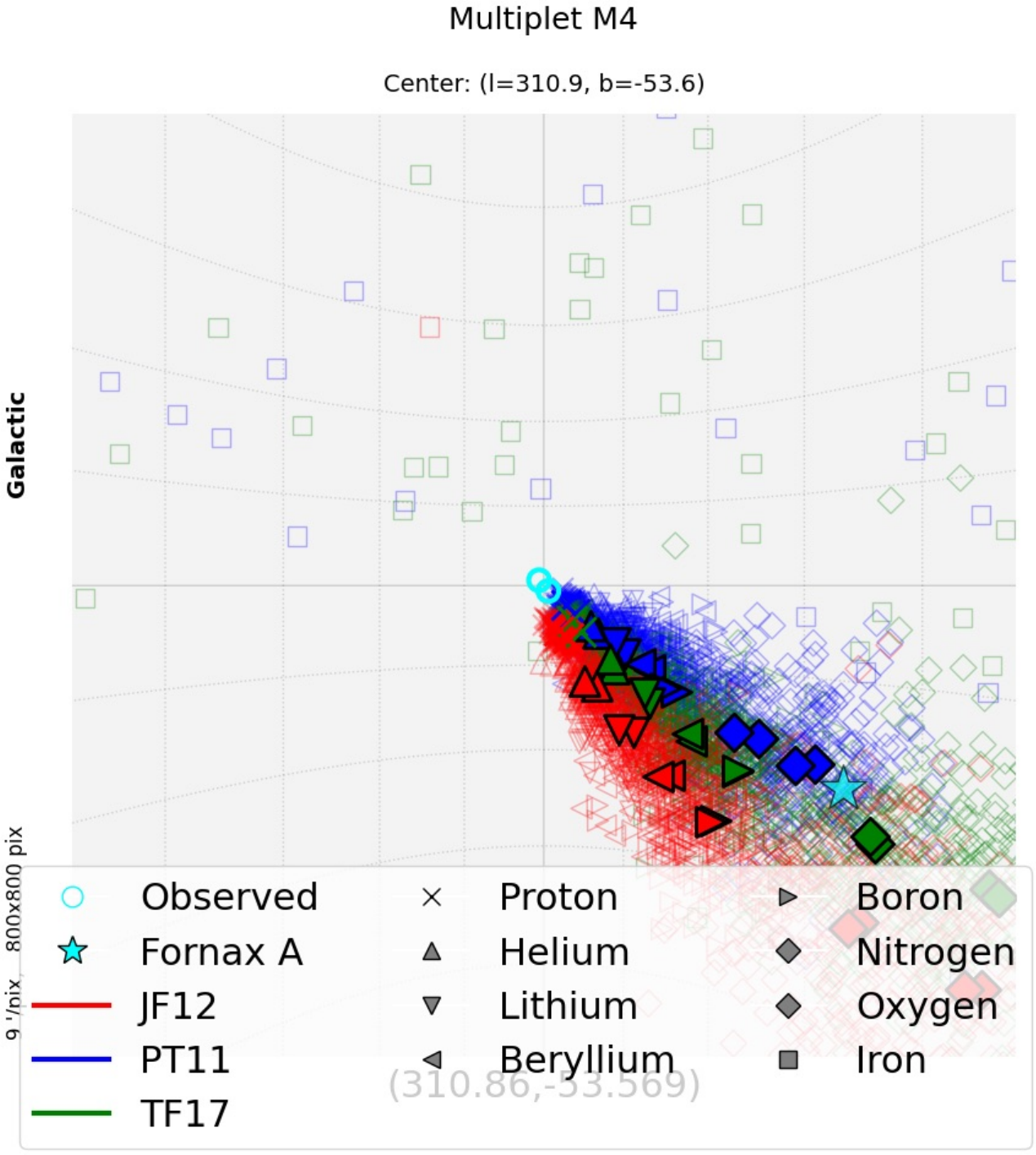}}
    \end{minipage}
    
    \caption{The backtracked positions of four doublets for different GMF models and nuclear compositions. Different colors correspond to different nuclei as indicated in the legend. The solid markers represent the median backtracking position for each nucleus, while the open, semi-transparent outlines show the 100 individual realizations. The cyan circles mark the arrival directions of the observed doublets, while the cyan star indicates the position of the closest source in the catalog. The background grid represents Galactic coordinates with a cell spacing of $10^\circ \times 10^\circ$. For all the multiplets see Appendix~\ref{app:all_multiplets}.}
    \label{fig:backtracking_grid}
\end{figure}

To quantify the preferred source and nuclear composition for each backtracked doublet, we applied the method from Section~\ref{sec:significance}. For each doublet, we identified the nucleus-source combination that maximizes the score, defined as the highest PDF value at the source coordinates. We evaluated the pre-trial chance probability ($P_{\text{Pre}}$) via Monte Carlo simulations, then corrected for the look-elsewhere effect by accounting for the 10 sources ($\sigma_{\text{Local}}$) tested. Hereafter we adopt JF12 as the default GMF model (see Appendix~\ref{app:doublets_systematics} for other models). Table~\ref{tab:individual_results} lists the full associations and their statistical significance, reflecting these trial factor penalties and the physical viability of these multiplets by calculating the survival rate ($f_{\text{surv}}$) for each constituent particle. We provide the MJD values for the events in the doublets to allow the community to diagnose the results.

\begin{table}[htbp]
\centering
\caption{Significance of individual multiplet-source-nucleus associations. The observed energies ($E$), the GMF deflection ($\delta_{\text{GMF}}$), and arrival times (Modified Julian Dates, MJD) of the events are listed, alongside the pre-trial probabilities ($P_{\text{pre}}$) and local significances ($\sigma_{\text{post}}$) of the associations. Simulated survival rates ($f_{\text{surv}}$) are calculated as upper limits using the observed energy.}
\label{tab:individual_results}
\setlength{\tabcolsep}{2pt}
\footnotesize
\begin{tabular}{lccccccccc}
\hline
\textbf{ID} & \textbf{Source} & \textbf{Iso} & \textbf{MJD} & \textbf{E [EeV]} & \textbf{$\delta_{\text{GMF}}$ [$^\circ$]} & \textbf{$P_{\text{pre}}$} & \textbf{$\sigma_{\text{pre}}$} & \textbf{$\sigma_{\text{post}}$} & \textbf{$f_{\text{surv}}$} \\
\hline
1 & Fornax A & He & 54182.92 / 54188.76 & 36.6 / 53.4 & 11.0 & $4.88 \times 10^{-4}$ & 3.3 & 2.6 & 0.03 / 0.00 \\
8 & Fornax A & He & 55775.52 / 55787.49 & 42.3 / 33.0 & 12.9 & $6.10 \times 10^{-4}$ & 3.2 & 2.5 & 0.00 / 0.11 \\
7 & Fornax A & Be & 55751.51 / 55757.47 & 43.4 / 45.5 & 22.4 & $1.59 \times 10^{-3}$ & 3.0 & 2.1 & 0.00 / 0.00 \\
9 & CGCG 114-025 & He & 55807.81 / 55814.09 & 33.1 / 87.5 & 28.7 & $4.15 \times 10^{-3}$ & 2.6 & 1.7 & 0.00 / 0.00 \\
15 & NGC 5090 & Li & 56674.37 / 56683.07 & 38.9 / 35.9 & 10.4 & $8.54 \times 10^{-3}$ & 2.4 & 1.4 & 0.00 / 0.00 \\
27 & Fornax A & N & 59048.73 / 59060.32 & 47.2 / 33.1 & 26.1 & $2.53 \times 10^{-2}$ & 2.0 & 0.8 & 0.34 / 0.65 \\
24 & NGC 5090 & O & 58371.86 / 58382.83 & 39.9 / 67.4 & 15.2 & $2.62 \times 10^{-2}$ & 1.9 & 0.7 & 0.38 / 0.01 \\
4 & Fornax A & N & 54602.48 / 54609.33 & 41.4 / 40.6 & 45.6 & $2.93 \times 10^{-2}$ & 1.9 & 0.7 & 0.48 / 0.52 \\
17 & Fornax A & O & 56807.00 / 56820.70 & 36.8 / 39.8 & 32.5 & $3.17 \times 10^{-2}$ & 1.9 & 0.6 & 0.71 / 0.65 \\
25 & NGC 0315 & N & 58548.59 / 58558.54 & 33.0 / 38.5 & 90.7 & $3.82 \times 10^{-2}$ & 1.8 & 0.5 & 0.26 / 0.15 \\
5 & M87 & Be & 54607.93 / 54608.11 & 34.7 / 62.5 & 26.5 & $4.19 \times 10^{-2}$ & 1.7 & 0.4 & 0.00 / 0.00 \\
23 & Fornax A & N & 58048.66 / 58063.20 & 33.0 / 32.6 & 34.5 & $4.75 \times 10^{-2}$ & 1.7 & 0.3 & 0.67 / 0.67 \\
0 & NGC 0315 & N & 53732.80 / 53733.82 & 32.9 / 38.0 & 93.8 & $5.95 \times 10^{-2}$ & 1.6 & 0.1 & 0.24 / 0.18 \\
18 & NGC 5090 & O & 56921.66 / 56930.83 & 88.9 / 61.6 & 15.6 & $6.52 \times 10^{-2}$ & 1.5 & 0.0 & 0.00 / 0.05 \\
26 & UGC 7360 & Fe & 59011.71 / 59012.97 & 155.2 / 62.2 & 50.1 & $7.31 \times 10^{-2}$ & 1.4 & 0.0 & 0.20 / 0.64 \\
16 & M87 & Fe & 56806.93 / 56812.70 & 45.9 / 52.5 & 26.9 & $7.52 \times 10^{-2}$ & 1.4 & 0.0 & 0.70 / 0.67 \\
3 & 2MASX J1254 & O & 54515.21 / 54521.70 & 33.0 / 41.2 & 50.1 & $8.89 \times 10^{-2}$ & 1.4 & 0.0 & 0.43 / 0.35 \\
12 & M87 & Fe & 56046.75 / 56054.00 & 35.8 / 39.1 & 42.2 & $1.09 \times 10^{-1}$ & 1.2 & 0.0 & 0.81 / 0.79 \\
6 & 2MASX J1254 & O & 55462.53 / 55463.34 & 38.5 / 33.7 & 24.0 & $1.14 \times 10^{-1}$ & 1.2 & 0.0 & 0.34 / 0.41 \\
10 & M87 & Fe & 55891.38 / 55895.19 & 39.5 / 56.7 & 60.9 & $1.30 \times 10^{-1}$ & 1.1 & 0.0 & 0.78 / 0.64 \\
20 & M87 & Fe & 57191.79 / 57195.90 & 35.3 / 57.9 & 10.2 & $1.63 \times 10^{-1}$ & 1.0 & 0.0 & 0.82 / 0.60 \\
14 & M87 & Fe & 56489.51 / 56492.55 & 39.6 / 32.2 & 74.5 & $1.70 \times 10^{-1}$ & 0.9 & 0.0 & 0.79 / 0.84 \\
11 & M87 & O & 55914.59 / 55926.55 & 33.7 / 33.8 & 14.0 & $1.00$ & 0.0 & 0.0 & 0.76 / 0.76 \\
21 & Fornax A & O & 57200.43 / 57202.21 & 80.3 / 72.8 & 29.3 & $1.00$ & 0.0 & 0.0 & 0.01 / 0.04 \\
\hline
\end{tabular}
\end{table}

Individual doublet associations are inherently weak, particularly for heavier nuclei which exhibit larger magnetic deflections and the cloud of the reconstructed trajectories become more smeared out. While we do not assume every doublet must correspond to a cataloged source, the recurrent association of multiple doublets with Fornax A is striking. This observation motivates the stacked analysis for all doublets associated with a given source (Section~\ref{sec:stacking}). We note that Centaurus A and NGC 5090 are spatially degenerate given the angular resolution; however, our algorithm consistently assigns a higher score to NGC 5090, which is therefore retained as the primary association. Table~\ref{tab:stacked_results} presents the post-trial significance for each stacked association, incorporating the appropriate trial factors to account for the scan over multiple sources and isotopes.

\begin{table}[htbp]
    \centering
    \caption{Significance of stacked source associations. The `Multiplets' column lists all doublet IDs contributing to the stacked score for each source. $P_{\text{pre}}$ denotes the pre-trial chance probability, and $\sigma_{\text{post}}$ represents the post-trial conditional significance.}
    \label{tab:stacked_results}
    
    % Resize table to fit text width
    \resizebox{\textwidth}{!}{%
    \begin{tabular}{llccc}
        \hline
        \textbf{Source} & \textbf{Multiplets} & \textbf{$P_{\text{pre}}$} & \textbf{$\sigma_{\text{pre}}$} & \textbf{$\sigma_{\text{post}}$} \\
        \hline
        Fornax A & M1, M17, M21, M23, M27, M4, M7, M8 & $4.0 \times 10^{-9}$ & 5.4 & 4.5 \\
        NGC 5090 & M15, M18, M24 & $1.7 \times 10^{-5}$ & 3.6 & 2.2 \\
        NGC 0315 & M0, M25 & $9.2 \times 10^{-5}$ & 3.1 & 1.4 \\
        M87 & M10, M11, M12, M14, M16, M20, M5 & $5.1 \times 10^{-3}$ & 1.6 & 0.0 \\
        2MASX J12543570 & M3, M6 & $1.3 \times 10^{-2}$ & 1.1 & 0.0 \\
        \hline
    \end{tabular}%
    }
\end{table}

The joint analysis identifies Fornax A as the primary source candidate, yielding a conditional stacked post-trial 4.5$\sigma$ for its set of associated multiplets. NGC\,5090 ranks second ($2.4\sigma$), based on an association with only three doublets. Conversely, Cen\,A shows a negligible association, despite the presence of a known hotspot in the overall dataset (see discussions below). Finally, evaluating the cumulative probability that the catalog of closest radio galaxies explains the detected doublets yields a total significance of $5.9\sigma$. When accounting for the chance probability of randomly finding 28 multiplets in the dataset, this significance is marginally reduced to $5.8\sigma$. We emphasize that this value is a conditional metric—calculated by fixing the specific event-to-source associations identified in the multiplet search—rather than an absolute, blind detection probability. Consequently, while it cannot be interpreted as a formal $5\sigma$ discovery, it serves as a robust statistical rejection of the null hypothesis for the ensemble. We discuss the implications of these results in the next section and analyze the impact of the choice of parameters (GMF model and time window) in Appendix~\ref{app:doublets_systematics}.

\section{Discussions}
\label{sect:discussions}

The method presented in this work relies on the identification and analysis of multiplets. In the dataset, only doublets were found, with no higher-order multiplets detected. While finding 28 doublets is not statistically rare—as shown in Fig.~\ref{fig:multiplet_sig} (left)—two key indicators suggest that at least a subset of these doublets is physical, originating from common sources rather than chance coincidence. First, the spatial clustering of doublets M1 and M8, which arrive within 3$^\circ$ of one another, is unlikely to occur by chance ($p = 1.4 \times 10^{-3}$ or 3.0$\sigma$ significance). Second, the alignment of several doublets follows the expected deflection path of charged particles through the GMF, a trend that is particularly pronounced for events arriving from the direction of Fornax A. Although individual associations are heavily penalized by trial factors, the stacked analysis of doublets associated with the same source yields a conditional stacked post-trial significance of $4.5\sigma$ for Fornax A (Table~\ref{tab:stacked_results}). It is important to emphasize that this significance is not a typical detection against a background; rather, it represents the probability of such a correlation occurring by chance given the anisotropic distribution of local matter. The result demonstrates that the observed signal is significantly more concentrated than what would be expected from the geometric clustering of the local Large Scale Structure, providing a strong indication that Fornax A is the physical source of these associated events.

The absence of a strong signal from Centaurus A is initially puzzling but physically explainable. Because Fornax A lies at a high Galactic latitude ($b \approx -56^\circ$), the GMF along its line of sight is significantly weaker and less turbulent than toward the Centaurus A region ($b \approx +19^\circ$). Consequently, the backtracked event clouds for Fornax A are more compact and well aligned, resulting in sharper 2D Gaussian fits and higher association likelihood scores. Furthermore, our algorithm identifies NGC\,5090 as the preferred source over Centaurus A for certain multiplets, despite the two sources being separated by only $\sim$1$^\circ$. This separation is comparable to the angular resolution of the Pierre Auger Observatory. Assuming events with energies above 10\,EeV trigger at least 6 stations, the angular resolution is typically better than 0.9$^\circ$ \cite{Auger_ang_resolution}. However, as shown in Appendix~\ref{app:egal}, when accounting for the expected deflection by the turbulent EGMF, we cannot distinguish Centaurus A from NGC\,5090 as potential sources for these doublets.

To better understand the physical origin of the associations, we evaluated the survivability of the observed nuclei. Associations involving light nuclei (Li, Be) exhibit negligible survival rates ($f_{\text{surv}} \approx 0$), even for nearby sources such as Fornax A. Helium candidates similarly show low survival probabilities ($\le 4\%$), with the notable exception of the lower-energy event ($33.0$~EeV) in doublet 8, which reaches $10\%$. In contrast, heavier nuclei, e.g., Nitrogen candidates associated with Fornax A and NGC 0315, generally show significant survival rates ($16\% - 68\%$). Iron candidates consistently display high survival probabilities ($\ge 62\%$), despite always showing negligible post-trial significances ($\sigma_{\text{post}} = 0.0$) in the correlations. The negligible survival rates for the most significant associations strongly suggest that these light-nucleus events are not primary cosmic rays but rather "daughter" nuclei—by-products of photo-disintegration and other interaction processes occurring during propagation from the source to Earth. However, an independent composition-measurement ($X_{max}$) that could work as a test for the nucleus-doublet association is not available in the dataset. On the other hand, if the true nuclear species of the events in a doublet differs from our assignment, the true source would have to be an unassociated UHECR accelerator lying in close angular proximity to the analyzed candidate. Given the scarcity of local UHECR sources, such a chance spatial coincidence is improbable. Furthermore, if these events were actually highly deflected heavy nuclei from angular distant sources, the probability that multiple independent trajectories would accidentally align to form coherent multiplets around Fornax A is rather unlikely.

When a parent nucleus fragments, the daughter particles retain approximately the primary's Lorentz factor, sharing the total energy proportionally to their masses ($E_{\text{daughter}} \approx (A_{\text{daughter}}/A_{\text{parent}}) E_{\text{parent}}$). Consequently, the rigidity of a fragment relative to its parent depends strictly on their mass-to-charge ratios: $R_{\text{daughter}} = R_{\text{parent}} [(A/Z)_{\text{daughter}} / (A/Z)_{\text{parent}}]$. Because the photo-disintegration cascade is dominated by asymmetric nucleon loss, rigidity diverges at most intermediate decay steps. It is only preserved ($\Delta R = 0$) in specific splits, such as  \(^{16}\mathrm{O} \to {}^{4}\mathrm{He} + {}^{12}\mathrm{C}\). Importantly, however, even in perfectly rigidity-matched events, the secondary nuclei comprising an observed doublet do not originate from the exact same parent particle. During extragalactic propagation, photo-disintegration via interactions with the cosmic microwave and extragalactic background light~\cite{Puget1976} impart a transverse momentum of up to $p_T \approx 40 - 150$~MeV/c to the ejected fragments. While the extreme Lorentz boost yields a microscopic divergence angle ($\theta \approx p_T c / E \sim 10^{-11 \circ}$)—allowing propagation codes~\cite{AlvesBatista2016} to safely treat these kinematics as strictly collinear—this small divergence translates into a physical lateral separation on the order of Astronomical Units after traversing typical extragalactic distances ($D \sim 10$~Mpc). This extragalactic separation scale is derived by scaling the geometric kinematic framework originally established for solar photodisintegration by Gerasimova and Zatsepin~\cite{Gerasimova1960}. Because the Pierre Auger Observatory spans approximately 50~km, the probability of detecting multiple fragments from a single extragalactic photo-disintegration event is negligible. Therefore, the members of a detected 15-day doublet must necessarily be independent particles—either independent primaries or secondary fragments from entirely different parent nuclei—that coincidentally possess exactly matched rigidities and aligned trajectories.

To evaluate the identical-rigidity requirement ($\Delta R \approx 0$) for the events to have taken the same path across the GMF, we analyze the reconstructed energies of the events in the doublets. If two independent particles possess identical rigidities and belong to the same assigned nuclear species ($Z_1 = Z_2$), as assumed in the methodology, their energies must be equal ($E_1 = E_2$). Assuming a statistical energy resolution of $\sigma_E/E \approx 14\%$ \cite{dataset_auger}, we apply this test to the magnetic rigidities ($R = E/Z$) of the doublets associated with Fornax A. For Doublet 1 (He), with $R_1 = 18.30$~EV and $R_2 = 26.70$~EV, the combined uncertainty on the difference is:

\begin{equation}\sigma_{\Delta R} = \sqrt{(0.14 \times 18.30)^2 + (0.14 \times 26.70)^2} \approx 4.53~\text{EV.}\end{equation}

The observed gap of $\Delta R = 8.40$~EV corresponds to a $1.85\sigma$ spread. Other Fornax A associations demonstrate varying internal consistency, with Doublets 8 (He) and 7 (Be) showing tight internal spreads of $1.24\sigma$ and $0.24\sigma$, respectively. These values indicate that the events from Doublets 7 and 8 may probe similar magnetic volumes, consistent with being emitted at similar times. However, this tight consistency varies across the dataset; for example, Doublet 9 (He), associated with CGCG 114-025, exhibits a much broader internal spread of $4.15\sigma$, suggesting either a more complex magnetic propagation path or a chance alignment.

\begin{figure}[t]
    \centering
    \includegraphics[width=\textwidth]{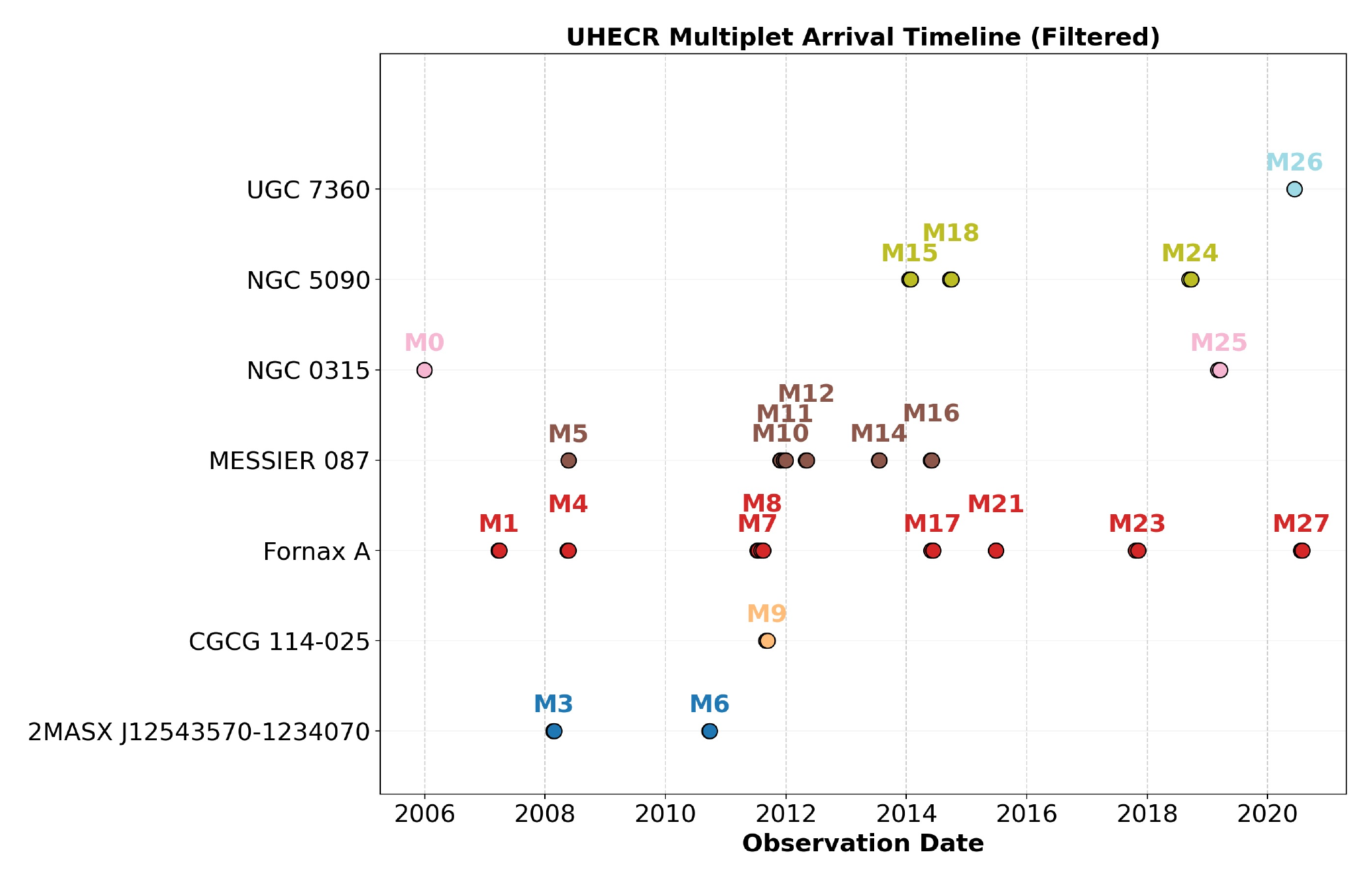}
    \caption{Arrival timeline of identified UHECR doublets (2004--2020). The vertical axis denotes the candidate source association, while each marker represents a doublet arrival. The steady distribution of events, particularly for Fornax~A, supports a long-term acceleration at the source.}
    \label{fig:timeline}
\end{figure}

Although we cannot affirm that the events within each doublet were emitted simultaneously, the steady detection of Fornax A doublets over time (Fig.~\ref{fig:timeline}) points to a long-lived accelerator. An isolated flare could theoretically produce a sequence of delayed detections; however, magnetic dispersion would lead to the highest-rigidity particles arriving first. Instead, the Fornax A multiplets appear intermittently throughout the 16-year window with randomly varying rigidities. This lack of chronological sorting suggests a source that continuously populates the propagation volume via persistent steady-state emission or recurring episodic flares. This is consistent with the scenario proposed by Matthews et al. \cite{Matthews2019}, in which giant radio lobes act as UHECR reservoirs. In this model, the particles we observe today were likely accelerated during a more powerful, possibly FRII-like, jet outburst in the source's history. The current arrival of doublets would thus represent the 'slow leaking' of these heavy nuclei from the extended radio lobes into the extragalactic medium over secular timescales.

As detailed in Appendix~\ref{app:time_delay}, the observed angular separations ($\theta_{\text{obs}} \sim 1^\circ$--$3^\circ$) vastly exceed the theoretical deflections required to explain the short time delays ($\theta_{\text{req}} \sim 0.001^\circ$). Therefore, large-scale magnetic path differences cannot explain the spatial spreading without introducing centuries-long temporal dispersions (Table~\ref{tab:detailed_timing}). Rather, the residual angular deviation between backtracked multiplets and their candidate sources—along with their internal separation, $\theta_{\text{obs}}$—results from an interplay of physical and experimental effects. Experimentally, these stem from the Pierre Auger Observatory's reconstruction uncertainties in arrival direction ($\sigma_{\text{exp}} \approx 1^\circ$) and energy, combined with the inherent degeneracy of event-by-event mass assignment. Physically, minor isotopic rigidity shifts ($\Delta R$) between fragments induce small-scale divergence, which is further amplified by continuous energy losses and limitations in current GMF/EGMF models.

Since our backtracking simulations focus exclusively on the GMF, it is necessary to estimate the systematic uncertainty introduced by neglecting the EGMF. To this end, we performed analytic simulations of the propagation of the associated nuclei, as detailed in Appendix~\ref{app:egal}. By comparing the predicted mean EGMF deflections ($\delta_\mathrm{mean}$ in Table~\ref{tab:egmf_results}) with the GMF-induced shifts ($\delta_\mathrm{GMF}$ in Table~\ref{tab:individual_results}), we find that the extragalactic contribution is approximately 20\% of the Galactic deflection for the most significant associations. While this contribution is not strictly negligible, estimating it requires assumptions about the EGMF strength and coherence length, which remain highly uncertain. Therefore, to avoid introducing additional model-dependent systematics, we consider the GMF-only approximation sufficient for the scope of this work.

We emphasize that we do not claim a definitive physical association for every individual multiplet identified. As shown in the backtracking maps (Appendix~\ref{app:all_multiplets}), several candidates remain visually offset from their nearest potential sources, and their individual significance levels (Table~\ref{tab:individual_results}) are consistent with the local matter density distribution. Nevertheless, the collective signal is robust. We observe a striking spatial conditional correlation of backtracked doublets with Fornax~A at the $4.5\sigma$ conditional (post-trial) level, reinforcing the $5.8\sigma$ conditional  association with the radio galaxy catalog. The significance quoted is conditional on the associated set of multiplets (Table~\ref{tab:individual_results}) and should therefore not be interpreted as a global significance of blind-search. The results suggest that while the analysis likely includes chance coincidences, the tight 15-day window effectively suppresses chance coincidence events, as confirmed by the 30-day window analysis in Appendix~\ref{app:doublets_systematics}.

The steady arrival of the doublets suggests that objects like Fornax~A function as long-term accelerators of high-energy particles. This is consistent with models where UHECRs are accelerated within the mildly relativistic backflows of extended radio lobes and subsequently escape into the extragalactic medium over secular timescales~\cite{Matthews2018, Matthews2019}. While the intense radiation fields of a central AGN often cause catastrophic photo-disintegration before heavy nuclei can escape the host, the lower photon densities within the giant lobes allow for the survival and gradual escape of the observed intermediate and heavy compositions. The observed temporal distribution of the doublets implies that this acceleration and escape process is characterized by a persistent emission history. While our data favor a long-term acceleration phase, the 16-year observational baseline of the Pierre Auger Observatory is orders of magnitude shorter than typical AGN duty cycles and magnetic delay times. Consequently, this exposure is insufficient to test for the existence of quiescent periods or transient flares.

Recent hydrodynamic simulations demonstrate that UHECRs are efficiently accelerated by diffusive shock acceleration in these supersonic backflows \cite{Matthews2019}. We can verify its capability to accelerate heavy primaries using the Hillas criterion, $E_{\text{max}} \simeq 0.9 \text{ EeV} \cdot Z (B/\mu\text{G}) (v_s/c) (R/\text{kpc})$ \cite{Matthews2019}. For the backflow shocks in radio galaxy lobes, assuming a characteristic shock size $R \sim 2$~kpc, a shock velocity $v_s \sim 0.2 c$, and an amplified magnetic field $B \sim 140~\mu$G \cite{Matthews2019}, the maximum attainable energy for an Iron primary ($Z=26$) is $E_{\text{max}} \approx 1300$~EeV. This theoretical maximum comfortably exceeds the primary energies ($\sim 500$--$800$~EeV) required to produce the $\sim 40$--$50$~EeV secondary fragments observed in the dataset. Achieving such energies requires a 'Goldilocks' situation where a jet power of $\sim 10^{44}$ erg s$^{-1}$ is channeled through a strong shock without reaching excessive velocities \cite{Matthews2019}. The association of 8 multiplets with Fornax A provides empirical evidence that these specific physical conditions for backflow-driven acceleration were likely met during its peak activity phases, enabling the production of the heavy primary nuclei inferred in our analysis. Our findings therefore support a scenario in which radio galaxies act as powerful accelerators of heavy UHECRs, injected over long timescales. This signature is detected at Earth primarily as lighter secondary nuclei produced through disintegration processes during propagation.

\section{Conclusions}
\label{sect:conclusions}

We analyzed 16 years of public UHECR data from the Pierre Auger Observatory \cite{dataset_auger} to identify multiplets—events arriving within a pre-defined spatiotemporal window of $3^\circ$ and 15 days of one another. While the tight spatial window acts as a rigidity filter, selecting events that likely follow similar paths along the GMF, the tight temporal window serves to suppress the rate of finding doublets by pure chance in the null-hypothesis scenario in which sources are distributed as the local matter in the universe. 

We identified 28 doublets and employed CRPropa 3 \cite{crpropa3} to backtrack their trajectories using three GMF models \cite{JF12,PT11,TF17} and eight nuclear species, testing associations against a catalog of the ten nearest bright radio galaxies \cite{vanVelzen_2012}. Using a Monte Carlo approach to assess the significance of the preferred nucleus-source pairings. We identified eight doublets associated with Fornax~A, resulting in a conditional stacked post-trial significance of $4.5\sigma$ (with nuclear species fixed to their individual best fits). According to the likelihood estimate, the most significant of these associations are secondary Helium and Beryllium. Conversely, the Centaurus~A--NGC 5090 region yielded a lower $2.4\sigma$ post-trial association from three doublets. We attribute this to the stronger and more turbulent GMF at the Centaurus~A latitude compared to Fornax~A, resulting in greater angular deflections, smearing out a more concentrated signal. This reduces the significance of associations under our methodology, which is better suited to the more concentrated backtracking characteristic of lighter nuclei. Testing the population of local radio galaxies as a whole yields a stacked post-trial significance of $5.8\sigma$, conditioned to the specific nucleus-source associations identified in the individual backtracking analysis. Consequently, while it cannot be interpreted as a formal $5\sigma$ discovery, it serves as a robust statistical rejection of the null hypothesis, in which the distribution of UHECR follows the anisotropic distribution of local matter \cite{Huchra2012}. While the
hypothesis was framed around the catalog as a population, the signal is undeniably dominated by Fornax A specifically.

A surviving ratio estimate indicates that most of the light-nucleus doublets are by-products of disintegration processes along the way. The fragments from photo-disintegration cannot originate from the same parent nucleus due to the transverse momentum the by-products acquire and the resulting rigidity-dependent delays. While some doublets have events with compatible rigidities, suggesting they could have been emitted at similar times from the same source, this is not a universal feature of the multiplet sample. Consequently, the tight 15-day arrival window does not necessarily capture single-epoch coincidences but serves to keep the probability of finding doublets by chance low.

The Fornax A associated doublets exhibit varying degrees of rigidity consistency. While some pairs possess matching rigidities compatible with simultaneous emission, others display significant differences that imply emission times separated by centuries to millennia, pointing to a long-term accelerator. Given that the 16-year observational baseline of the Pierre Auger Observatory is negligible compared to typical AGN duty cycles and magnetic delays, these results are best reconciled with the 'leaking reservoir' model \cite{Matthews2019}. In this scenario, heavy nuclei ($Z > 3$) were accelerated within mildly relativistic backflows during intense jet outbursts—possibly during a more powerful FRII-like phase in the galaxy's history—with the giant radio lobes now functioning as long-term storage sites that slowly release these particles into the extragalactic medium. Our methodology functions as a strict kinematic filter, isolating high-rigidity particles; while this targeted sample is consequently unsuitable for estimating global astrophysical flux levels or mass composition, it provides precise spatial insights. Collectively, our findings reinforce local radio galaxies as a primary source population for UHECRs, with Fornax A emerging as a prominent contributing accelerator within this population.
\newpage

\appendix
\section{Additional results - varying initial parameters}
\label{app:doublets_systematics}

In this section, we extend our main analysis by evaluating nucleus-source associations and their corresponding significances using the deflections predicted by the PT11 and TF17 GMF models. Additionally, we repeat the multiplet search and the complete analysis pipeline for two alternative temporal windows (7 and 30 days).

\subsection{Varying the GMF model}

We maintain the baseline configuration (15-day temporal window, $3^\circ$ spatial separation, and the original 28 multiplets) to evaluate the impact of the GMF model. Replacing the JF12 model with the PT11 and TF17 models yields the individual and stacked association significances presented in Tables~\ref{tab:PT11_results} and \ref{tab:PT11_results_stacked} (for PT11) and Tables~\ref{tab:TF17_results} and \ref{tab:TF17_results_stacked} (for TF17). The post-trial stacked conditional significance of the radio galaxy catalog associated with these multiplets is $>$6.4$\sigma$ for the PT11 model (requiring $10^{10}$ Monte Carlo simulations of the source positions) and $6.1\sigma$ for the TF17 model. These drop to $>$6.3$\sigma$\footnote{The PT11 analysis has not delivered any simulated catalog that fits better the data than the radio catalog, hence a lower-limit on the significance.} and 6.2$\sigma$ for the PT11 model and TF17 model, respectively, after penalizing for the doublet accidental probabilities. Therefore, our results for Fornax\,A and the overall catalog are robust to the choice of GMF model, despite significant differences in the backtracked trajectories for several multiplets associated to other sources, as shown in Appendix~\ref{app:all_multiplets}. While the total count of associated multiplets varies between models (ranging from 7 in JF12 to 12 in TF17), the identity of Fornax A as the primary contributing source remains consistent across all tested geometries.

\begin{table}[htbp]
\centering
\caption{Significance of individual multiplet-source-nucleus associations sorted by significance for the PT11 GMF model. Arrival times (MJD), observed energies ($E$), and the GMF deflection ($\delta_{\text{GMF}}$) are included for each doublet.}
\label{tab:PT11_results}
\setlength{\tabcolsep}{3pt}
\footnotesize
\begin{tabular}{lcccccccc}
\hline
\textbf{ID} & \textbf{Source} & \textbf{Iso} & \textbf{MJD} & \textbf{E [EeV]} & \textbf{$\delta_{\text{GMF}}$ [$^\circ$]} & \textbf{$P_{\text{pre}}$} & \textbf{$\sigma_{\text{pre}}$} & \textbf{$\sigma_{\text{post}}$} \\
\hline
8 & Fornax A & He & 55775.5, 55787.5 & 42.3, 33.0 & 11.37 & $1.83 \times 10^{-4}$ & 3.6 & 2.9 \\
7 & Fornax A & B & 55751.5, 55757.5 & 43.4, 45.5 & 22.83 & $1.22 \times 10^{-3}$ & 3.0 & 2.2 \\
1 & Fornax A & He & 54182.9, 54188.8 & 36.6, 53.4 & 9.53 & $1.95 \times 10^{-3}$ & 2.9 & 2.1 \\
5 & M87 & Li & 54607.9, 54608.1 & 34.7, 62.5 & 8.43 & $2.74 \times 10^{-3}$ & 2.8 & 1.9 \\
4 & Fornax A & O & 54602.5, 54609.3 & 41.4, 40.6 & 35.11 & $5.18 \times 10^{-3}$ & 2.6 & 1.6 \\
21 & Fornax A & O & 57200.4, 57202.2 & 80.3, 72.8 & 22.79 & $1.49 \times 10^{-2}$ & 2.2 & 1.1 \\
23 & Fornax A & B & 58048.7, 58063.2 & 33.0, 32.6 & 29.29 & $3.51 \times 10^{-2}$ & 1.8 & 0.5 \\
27 & Fornax A & O & 59048.7, 59060.3 & 47.2, 33.1 & 36.33 & $3.82 \times 10^{-2}$ & 1.8 & 0.5 \\
15 & NGC 5090 & B & 56674.4, 56683.1 & 38.9, 35.9 & 11.74 & $4.60 \times 10^{-2}$ & 1.7 & 0.3 \\
17 & Fornax A & N & 56807.0, 56820.7 & 36.8, 39.8 & 27.04 & $5.90 \times 10^{-2}$ & 1.6 & 0.1 \\
24 & NGC 5090 & O & 58371.9, 58382.8 & 39.9, 67.4 & 8.43 & $6.65 \times 10^{-2}$ & 1.5 & 0.0 \\
18 & NGC 5090 & O & 56921.7, 56930.8 & 88.9, 61.6 & 11.45 & $1.18 \times 10^{-1}$ & 1.2 & 0.0 \\
19 & Fornax A & N & 57157.7, 57165.9 & 37.9, 107.1 & 18.69 & $1.34 \times 10^{-1}$ & 1.1 & 0.0 \\
13 & M87 & O & 56258.7, 56269.7 & 73.4, 38.5 & 20.52 & $1.00$ & 0.0 & 0.0 \\
22 & M87 & O & 57761.5, 57773.5 & 46.4, 42.5 & 23.38 & $1.00$ & 0.0 & 0.0 \\
\hline
\end{tabular}
\end{table}

\begin{table}[htbp]
    \centering
    \caption{Significance of stacked source associations. The `Multiplets' column lists all doublet IDs contributing to the stacked score for each source. $P_{\text{pre}}$ denotes the pre-trial chance probability, and $\sigma_{\text{post}}$ represents the post-trial significance.}
    \label{tab:PT11_results_stacked}
    
    % Resize table to fit text width
    \resizebox{\textwidth}{!}{%
    \begin{tabular}{llccc}
        \hline
        \textbf{Source} & \textbf{Multiplets} & \textbf{$P_{\text{pre}}$} & \textbf{$\sigma_{\text{pre}}$} & \textbf{$\sigma_{\text{post}}$} \\
        \hline
        Fornax A & M1, M17, M19, M21, M23, M27, M4, M7, M8 & $3.3 \times 10^{-9}$ & 5.4 & 4.5 \\
        M87 & M13, M22, M5 & $1.9 \times 10^{-4}$ & 2.9 & 1.1 \\
        NGC 5090 & M15, M18, M24 & $5.5 \times 10^{-3}$ & 1.6 & 0.0 \\
        \hline
    \end{tabular}%
    }
\end{table}

\begin{table}[htbp]
\centering
\caption{Significance of individual multiplet-source-nucleus associations sorted by significance for the TF17 GMF model. Arrival times (MJD), observed energies ($E$), and the GMF deflection ($\delta_{\text{GMF}}$) are included for each doublet.}
\label{tab:TF17_results}
\setlength{\tabcolsep}{3pt}
\footnotesize
\begin{tabular}{lcccccccc}
\hline
\textbf{ID} & \textbf{Source} & \textbf{Iso} & \textbf{MJD} & \textbf{E [EeV]} & \textbf{$\delta_{\text{GMF}}$ [$^\circ$]} & \textbf{$P_{\text{pre}}$} & \textbf{$\sigma_{\text{pre}}$} & \textbf{$\sigma_{\text{post}}$} \\
\hline
1 & Fornax A & He & 54182.9, 54188.8 & 36.6, 53.4 & 11.35 & $3.05 \times 10^{-4}$ & 3.4 & 2.7 \\
7 & Fornax A & Be & 55751.5, 55757.5 & 43.4, 45.5 & 22.47 & $4.27 \times 10^{-4}$ & 3.3 & 2.6 \\
8 & Fornax A & He & 55775.5, 55787.5 & 42.3, 33.0 & 13.02 & $5.49 \times 10^{-4}$ & 3.3 & 2.5 \\
15 & NGC 5090 & He & 56674.4, 56683.1 & 38.9, 35.9 & 4.82 & $1.16 \times 10^{-2}$ & 2.3 & 1.2 \\
5 & M87 & Li & 54607.9, 54608.1 & 34.7, 62.5 & 11.13 & $1.36 \times 10^{-2}$ & 2.2 & 1.1 \\
16 & Fornax A & B & 56806.9, 56812.7 & 45.9, 52.5 & 67.24 & $2.17 \times 10^{-2}$ & 2.0 & 0.8 \\
12 & Fornax A & N & 56046.8, 56054.0 & 35.8, 39.1 & 66.73 & $3.24 \times 10^{-2}$ & 1.9 & 0.6 \\
4 & Fornax A & B & 54602.5, 54609.3 & 41.4, 40.6 & 30.73 & $3.66 \times 10^{-2}$ & 1.8 & 0.5 \\
10 & Fornax A & O & 55891.4, 55895.2 & 39.5, 56.7 & 31.34 & $4.21 \times 10^{-2}$ & 1.7 & 0.4 \\
24 & NGC 5090 & O & 58371.9, 58382.8 & 39.9, 67.4 & 12.83 & $4.84 \times 10^{-2}$ & 1.7 & 0.3 \\
14 & Fornax A & Be & 56489.5, 56492.6 & 39.6, 32.2 & 37.70 & $5.20 \times 10^{-2}$ & 1.6 & 0.2 \\
27 & Fornax A & B & 59048.7, 59060.3 & 47.2, 33.1 & 28.91 & $6.79 \times 10^{-2}$ & 1.5 & 0.0 \\
20 & Fornax A & O & 57191.8, 57195.9 & 35.3, 57.9 & 82.05 & $6.97 \times 10^{-2}$ & 1.5 & 0.0 \\
18 & NGC 5090 & N & 56921.7, 56930.8 & 88.9, 61.6 & 17.81 & $7.56 \times 10^{-2}$ & 1.4 & 0.0 \\
6 & Fornax A & N & 55462.5, 55463.3 & 38.5, 33.7 & 37.53 & $7.87 \times 10^{-2}$ & 1.4 & 0.0 \\
17 & NGC 5090 & B & 56807.0, 56820.7 & 36.8, 39.8 & 31.03 & $1.21 \times 10^{-1}$ & 1.2 & 0.0 \\
19 & Fornax A & Li & 57157.7, 57165.9 & 37.9, 107.1 & 11.09 & $1.47 \times 10^{-1}$ & 1.1 & 0.0 \\
23 & Fornax A & B & 58048.7, 58063.2 & 33.0, 32.6 & 39.42 & $1.60 \times 10^{-1}$ & 1.0 & 0.0 \\
2 & NGC 5090 & O & 54470.0, 54474.0 & 37.9, 35.2 & 38.62 & $1.67 \times 10^{-1}$ & 1.0 & 0.0 \\
3 & NGC 5090 & Be & 54515.2, 54521.7 & 33.0, 41.2 & 24.04 & $1.00$ & 0.0 & 0.0 \\
\hline
\end{tabular}
\end{table}

\begin{table}[htbp]
    \centering
    \caption{Significance of stacked source associations for the TF17 GMF model. $P_{\text{pre}}$ denotes the pre-trial chance probability. Sources with only a single associated multiplet have been excluded.}
    \label{tab:TF17_results_stacked}
    
    % Resize table to fit text width
    \resizebox{\textwidth}{!}{%
    \begin{tabular}{l p{8cm} ccc}
        \hline
        \textbf{Source} & \textbf{Multiplets} & \textbf{$P_{\text{pre}}$} & \textbf{$\sigma_{\text{pre}}$} & \textbf{$\sigma_{\text{glo}}$} \\
        \hline
        Fornax A & M1, M10, M12, M14, M16, M19, M20, M23, M27, M4, M6, M7, M8 & $4.0 \times 10^{-11}$ & 6.1 & 5.4 \\
        NGC 5090 & M15, M17, M18, M2, M24, M3 & $1.8 \times 10^{-4}$ & 2.9 & 1.1 \\
        \hline
    \end{tabular}%
    }
\end{table}

\subsection{Varying the time window}

\begin{figure}[htbp]
    \centering
    \includegraphics[width=0.95\linewidth]{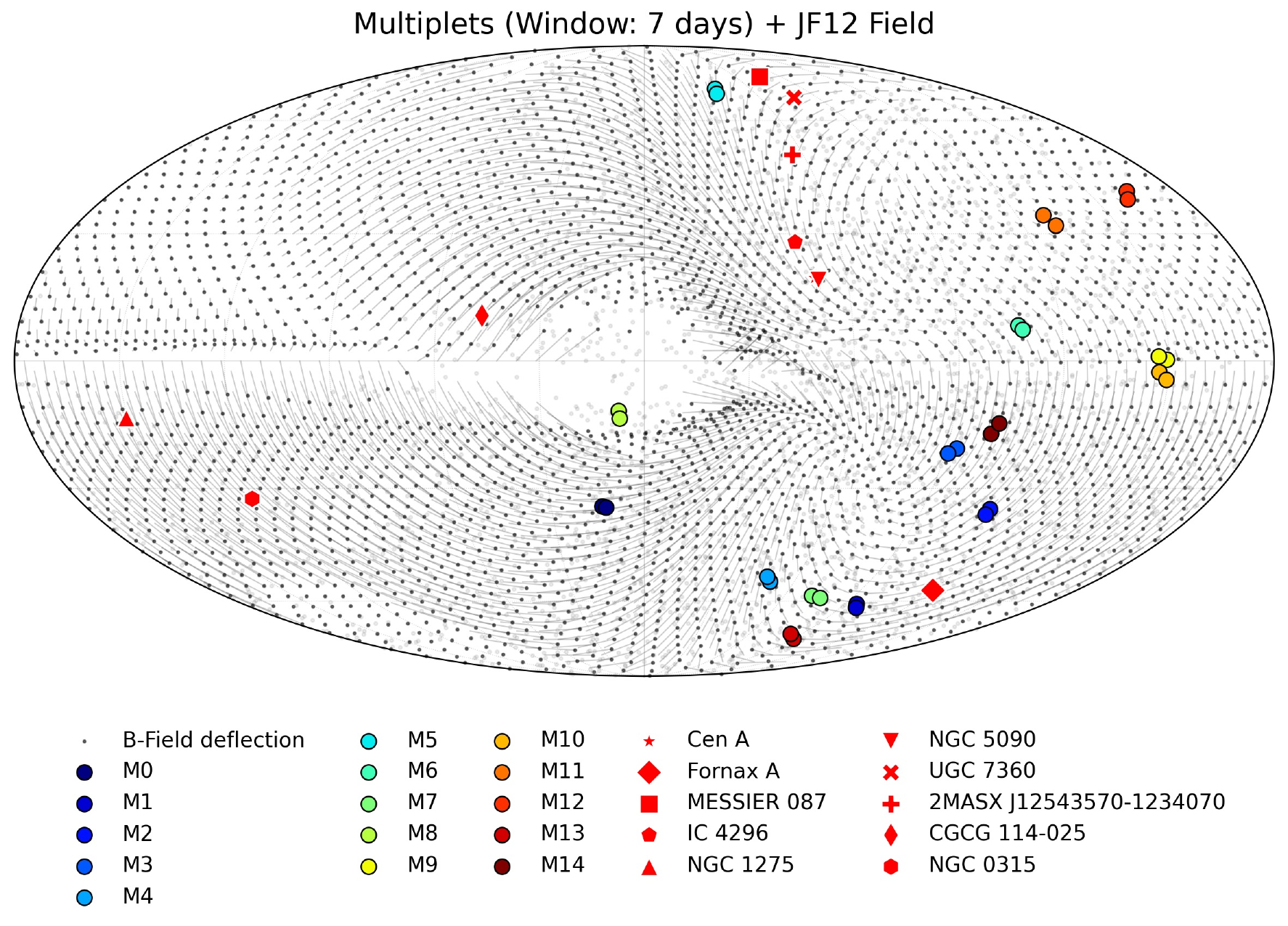}
    \caption{Sky map of the UHECR events in Galactic coordinates for the 7-day analysis. Analogous to Fig.~\ref{fig:multiplet_map}.
    \label{fig:multiplets_7}}
\end{figure}

\begin{figure}[htbp]
    \centering
    \includegraphics[width=0.95\linewidth]{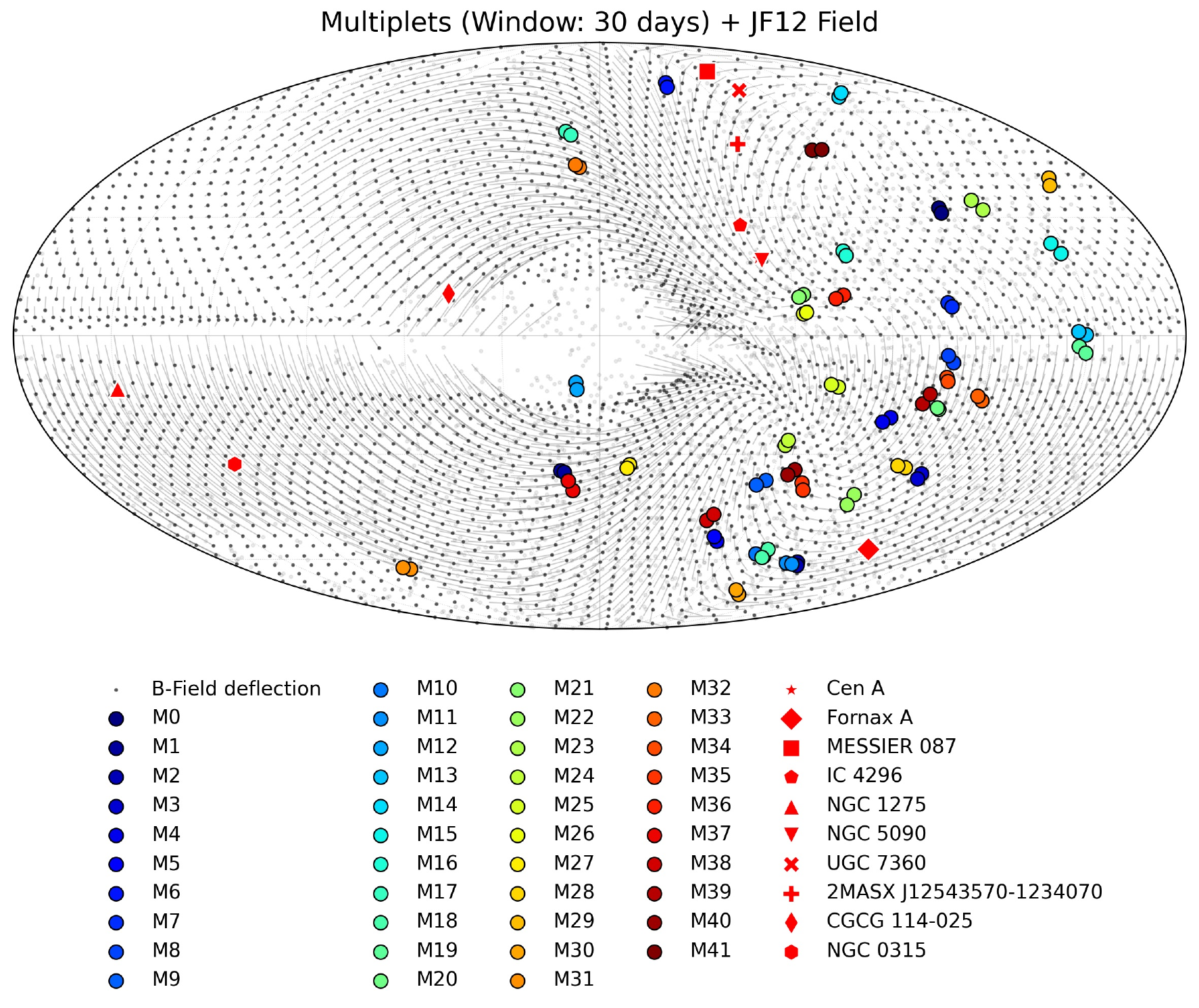}
    \caption{Sky map of the UHECR events in Galactic coordinates for the 30-day analysis. Analogous to Fig.~\ref{fig:multiplet_map}.
    \label{fig:multiplets_30}}
\end{figure}

We repeat the multiplet search and association analysis using alternative temporal windows of 7 and 30 days. This yields 15 unique multiplets for the 7-day window (Fig.~\ref{fig:multiplets_7}) and 42 for the 30-day window (Fig.~\ref{fig:multiplets_30}). Although many multiplets overlap with those in the 15-day baseline, their IDs are unique to each temporal configuration. The probability of finding exactly 15 (42) chance multiplets in the 7-day (30-day) analysis corresponds to a significance of $1.8\sigma$ ($1.0\sigma$). The stacked significance for the radio galaxy catalog is $4.8\sigma$ ($>6.4\sigma$) for these windows. After penalizing for the doublet accidental probabilities, the final post-trial conditional stacked significance values are $4.7\sigma$ ($>5.8\sigma$) for the 7-day (30-day) analysis, respectively. 

A narrow time window keeps the probability of chance coincidences relatively low, effectively compensating for the smaller doublet count and preserving a high association significance. Conversely, longer time windows yield more doublets but are heavily penalized by the increased number of chance coincidence pairings. Because the two particles in a doublet may follow distinct trajectories, extending the time window increases the probability of falsely associating unrelated events from distinct sources. This dynamic explains why correlated doublets are not typically significantly detected in time-integrated analyses.

Our a priori selection of the 15-day window balances signal retention with suppression of unrelated events in a doublet. As demonstrated by diagnostic checks, narrower windows (e.g., 7 days) artificially truncate the physical signal, leaving too few events for robust statistical analysis. Conversely, extended windows ($\ge 30$ days) rapidly dilute the true correlations by accumulating random chance alignments. Although the doublets method provides statistical evidence for the ensemble, it inherently cannot distinguish genuine astrophysical doublets from chance coincidences on an individual basis.

Tables~\ref{tab:7day_results} through \ref{tab:30day_results_stacked} detail the results for the 7-day and 30-day temporal windows. While the associations are robust at the 7-day threshold, their statistical significance increases substantially when expanding the window to 30 days. Most notably, the stacked association for Fornax A reaches a post-trial conditional significance of $5.6\sigma$ in the 30-day analysis. This increase is expected due to the larger number of associated multiplets captured within the extended time window. However, the probability of random  multiplets is correspondingly high, indicating that this result must be interpreted with caution.

As a final consistency check, the 2MRS catalog used for source randomization was replaced with a purely random isotropic distribution of potential sources. This sensitivity test yielded results nearly identical to the primary analysis: the global conditional correlation between the multiplets and the radio galaxy catalog reached the $5.5\sigma$ level, while the specific association of the eight doublets with Fornax A was confirmed at $4.5\sigma$ post-trial. The stability of these significance levels across different reference distributions demonstrates that the identified correlations are driven by the source positions themselves rather than artifacts of the underlying matter density model.

\begin{table}[htbp]
\centering
\caption{Significance of individual multiplet-source-nucleus associations sorted by significance for the 7-day time window. IDs refer to Fig.~\ref{fig:multiplets_7}}
\label{tab:7day_results}
\setlength{\tabcolsep}{3pt}
\footnotesize
\begin{tabular}{llccccccc}
\hline
\textbf{ID} & \textbf{Source} & \textbf{Iso} & \textbf{MJD} & \textbf{E [EeV]} & \textbf{$\delta_{\text{GMF}}$ [$^\circ$]} & \textbf{$P_{\text{pre}}$} & \textbf{$\sigma_{\text{pre}}$} & \textbf{$\sigma_{\text{glo}}$} \\
\hline
1 & Fornax A & He & 54182.9, 54188.8 & 36.6, 53.4 & 11.00 & $4.88 \times 10^{-4}$ & 3.3 & 2.6 \\
7 & Fornax A & Be & 55751.5, 55757.5 & 43.4, 45.5 & 22.36 & $1.59 \times 10^{-3}$ & 3.0 & 2.1 \\
8 & CGCG 114-025 & He & 55807.8, 55814.1 & 33.1, 87.5 & 28.70 & $4.15 \times 10^{-3}$ & 2.6 & 1.7 \\
4 & Fornax A & N & 54602.5, 54609.3 & 41.4, 40.6 & 45.61 & $2.93 \times 10^{-2}$ & 1.9 & 0.7 \\
5 & M87 & Be & 54607.9, 54608.1 & 34.7, 62.5 & 26.50 & $4.19 \times 10^{-2}$ & 1.7 & 0.4 \\
0 & NGC 0315 & N & 53732.8, 53733.8 & 32.9, 38.0 & 93.80 & $5.95 \times 10^{-2}$ & 1.6 & 0.1 \\
14 & UGC 7360 & Fe & 59011.7, 59013.0 & 155.2, 62.2 & 50.13 & $7.31 \times 10^{-2}$ & 1.4 & 0.0 \\
11 & M87 & Fe & 56806.9, 56812.7 & 45.9, 52.5 & 26.91 & $7.52 \times 10^{-2}$ & 1.4 & 0.0 \\
3 & 2MASX J1254 & O & 54515.2, 54521.7 & 33.0, 41.2 & 50.15 & $8.89 \times 10^{-2}$ & 1.4 & 0.0 \\
6 & 2MASX J1254 & O & 55462.5, 55463.3 & 38.5, 33.7 & 24.01 & $1.14 \times 10^{-1}$ & 1.2 & 0.0 \\
9 & M87 & Fe & 55891.4, 55895.2 & 39.5, 56.7 & 60.88 & $1.30 \times 10^{-1}$ & 1.1 & 0.0 \\
12 & M87 & Fe & 57191.8, 57195.9 & 35.3, 57.9 & 10.20 & $1.63 \times 10^{-1}$ & 1.0 & 0.0 \\
10 & M87 & Fe & 56489.5, 56492.6 & 39.6, 32.2 & 74.46 & $1.70 \times 10^{-1}$ & 0.9 & 0.0 \\
13 & Fornax A & O & 57200.4, 57202.2 & 80.3, 72.8 & 29.26 & $1.00$ & 0.0 & 0.0 \\
\hline
\end{tabular}
\end{table}

\begin{table}[htbp]
    \centering
    \caption{Significance of stacked source associations for the 7-day time window. $P_{\text{pre}}$ denotes the pre-trial chance probability. IDs refer to Fig.~\ref{fig:multiplets_7}. Sources with only a single associated multiplet have been excluded.}
    \label{tab:7day_results_stacked}
    
    % Resize table to fit text width
    %\resizebox{\textwidth}{!}{%
    \begin{tabular}{llccc}
        \hline
        \textbf{Source} & \textbf{Multiplets} & \textbf{$P_{\text{pre}}$} & \textbf{$\sigma_{\text{pre}}$} & \textbf{$\sigma_{\text{glo}}$} \\
        \hline
        Fornax A & M1, M13, M4, M7 & $7.0 \times 10^{-8}$ & 4.8 & 3.9 \\
        M87 & M10, M11, M12, M5, M9 & $6.0 \times 10^{-3}$ & 1.6 & 0.0 \\
        2MASX J1254 & M3, M6 & $1.3 \times 10^{-2}$ & 1.1 & 0.0 \\
        \hline
    \end{tabular}%
    %}
\end{table}

\begin{table}[htbp]
\centering
\caption{Significance of individual multiplet-source-nucleus associations sorted by significance for the 30-day time window. IDs refer to Fig.~\ref{fig:multiplets_30}}
\label{tab:30day_results}
\setlength{\tabcolsep}{3pt}
\footnotesize
\begin{tabular}{llccccccc}
\hline
\textbf{ID} & \textbf{Source} & \textbf{Iso} & \textbf{MJD} & \textbf{E [EeV]} & \textbf{$\delta_{\text{GMF}}$ [$^\circ$]} & \textbf{$P_{\text{pre}}$} & \textbf{$\sigma_{\text{pre}}$} & \textbf{$\sigma_{\text{glo}}$} \\
\hline
2 & Fornax A & He & 54182.9, 54188.8 & 36.6, 53.4 & 11.1 & $4.88 \times 10^{-4}$ & 3.3 & 2.6 \\
11 & Fornax A & He & 55775.5, 55787.5 & 42.3, 33.0 & 12.9 & $6.10 \times 10^{-4}$ & 3.2 & 2.5 \\
18 & Fornax A & Be & 56394.6, 56422.0 & 41.0, 41.1 & 23.4 & $8.54 \times 10^{-4}$ & 3.1 & 2.4 \\
10 & Fornax A & Be & 55751.5, 55757.5 & 43.4, 45.5 & 22.2 & $1.59 \times 10^{-3}$ & 3.0 & 2.1 \\
12 & CGCG 114-025 & He & 55807.8, 55814.1 & 33.1, 87.5 & 27.5 & $4.15 \times 10^{-3}$ & 2.6 & 1.7 \\
21 & NGC 5090 & Li & 56674.4, 56683.1 & 38.9, 35.9 & 9.5 & $8.54 \times 10^{-3}$ & 2.4 & 1.4 \\
26 & NGC 5090 & Li & 56958.6, 56975.6 & 33.6, 43.4 & 10.4 & $1.45 \times 10^{-2}$ & 2.2 & 1.1 \\
9 & Fornax A & N & 55696.9, 55712.4 & 32.2, 39.3 & 40.4 & $1.68 \times 10^{-2}$ & 2.1 & 1.0 \\
40 & Fornax A & N & 59048.7, 59060.3 & 47.2, 33.1 & 31.4 & $2.53 \times 10^{-2}$ & 2.0 & 0.8 \\
36 & NGC 5090 & O & 58371.9, 58382.8 & 39.9, 67.4 & 12.2 & $2.62 \times 10^{-2}$ & 1.9 & 0.7 \\
5 & Fornax A & N & 54602.5, 54609.3 & 41.4, 40.6 & 47.3 & $2.93 \times 10^{-2}$ & 1.9 & 0.7 \\
38 & Fornax A & O & 58672.3, 58694.0 & 65.8, 64.5 & 35.5 & $3.03 \times 10^{-2}$ & 1.9 & 0.6 \\
24 & Fornax A & O & 56807.0, 56820.7 & 36.8, 39.8 & 30.4 & $3.17 \times 10^{-2}$ & 1.9 & 0.6 \\
37 & NGC 0315 & N & 58548.6, 58558.5 & 33.0, 38.5 & 90.1 & $3.82 \times 10^{-2}$ & 1.8 & 0.5 \\
31 & NGC 1275 & N & 57264.2, 57290.2 & 41.3, 41.1 & 64.3 & $4.09 \times 10^{-2}$ & 1.7 & 0.4 \\
6 & M87 & Be & 54607.9, 54608.1 & 34.7, 62.5 & 25.9 & $4.19 \times 10^{-2}$ & 1.7 & 0.4 \\
35 & Fornax A & N & 58048.7, 58063.2 & 33.0, 32.6 & 33.8 & $4.75 \times 10^{-2}$ & 1.7 & 0.3 \\
27 & NGC 0315 & O & 56978.0, 56999.9 & 33.9, 43.8 & 95.0 & $5.03 \times 10^{-2}$ & 1.6 & 0.2 \\
41 & UGC 7360 & O & 59060.8, 59088.8 & 60.4, 33.3 & 13.9 & $5.17 \times 10^{-2}$ & 1.6 & 0.2 \\
1 & NGC 0315 & N & 53732.8, 53733.8 & 32.9, 38.0 & 106.5 & $5.95 \times 10^{-2}$ & 1.6 & 0.1 \\
25 & NGC 5090 & O & 56921.7, 56930.8 & 88.9, 61.6 & 15.7 & $6.52 \times 10^{-2}$ & 1.5 & 0.0 \\
16 & 2MASX J1254 & O & 56234.4, 56259.7 & 44.5, 39.3 & 13.9 & $6.85 \times 10^{-2}$ & 1.5 & 0.0 \\
0 & UGC 7360 & O & 53634.6, 53651.5 & 54.1, 32.3 & 9.9 & $7.50 \times 10^{-2}$ & 1.4 & 0.0 \\
23 & M87 & Fe & 56806.9, 56812.7 & 45.9, 52.5 & 27.5 & $7.52 \times 10^{-2}$ & 1.4 & 0.0 \\
8 & UGC 7360 & Fe & 55494.4, 55516.3 & 56.0, 65.2 & 26.8 & $7.73 \times 10^{-2}$ & 1.4 & 0.0 \\
34 & NGC 5090 & N & 57953.6, 57982.7 & 33.1, 68.8 & 35.1 & $8.21 \times 10^{-2}$ & 1.4 & 0.0 \\
4 & 2MASX J1254 & O & 54515.2, 54521.7 & 33.0, 41.2 & 48.5 & $8.89 \times 10^{-2}$ & 1.4 & 0.0 \\
7 & 2MASX J1254 & O & 55462.5, 55463.3 & 38.5, 33.7 & 21.5 & $1.14 \times 10^{-1}$ & 1.2 & 0.0 \\
19 & M87 & Fe & 56489.5, 56492.6 & 39.6, 32.2 & 72.6 & $1.70 \times 10^{-1}$ & 0.9 & 0.0 \\
\hline
\end{tabular}
\end{table}

\begin{table}[htbp]
    \centering
    \caption{Significance of stacked source associations for the 30-day time window. $P_{\text{pre}}$ denotes the pre-trial chance probability. IDs refer to Fig.~\ref{fig:multiplets_30}. Sources with only a single associated multiplet have been excluded.}
    \label{tab:30day_results_stacked}
    %\resizebox{\textwidth}{!}{%
    \begin{tabular}{lp{5cm}ccc}
        \hline
        \textbf{Source} & \textbf{Multiplets} & \textbf{$P_{\text{pre}}$} & \textbf{$\sigma_{\text{pre}}$}  & \textbf{$\sigma_{\text{glo}}$} \\
        \hline
        Fornax A & M2, M5, M9, M10, M11, M18, M22, M24, M30, M35, M38, M40 & $1.0 \times 10^{-11}$ & 6.4 & 5.7 \\
        NGC 5090 & M21, M25, M26, M34, M36 & $1.9 \times 10^{-6}$ & 4.1 & 3.0 \\
        NGC 0315 & M1, M27, M37 & $5.8 \times 10^{-6}$ & 3.8 & 2.6 \\
        UGC 7360 & M0, M8, M39, M41 & $3.5 \times 10^{-3}$ & 1.8 & 0.0 \\
        M87 & M6, M13, M14, M15, M19, M23, M29 & $5.1 \times 10^{-3}$ & 1.6 & 0.0 \\
        2MASX J1254 & M4, M7, M16 & $8.3 \times 10^{-3}$ & 1.4 & 0.0 \\
        \hline
    \end{tabular}%
    %}
\end{table}

\section{Backtracked trajectory of all doublets}
\label{app:all_multiplets}

%1784
\MultipletPage{27}{17}{23}{21}

\MultipletPage{9}{15}{24}{18}

\MultipletPage{25}{0}{26}{5}

\MultipletPage{16}{12}{10}{20}

\MultipletPage{14}{11}{6}{3}

\section{Estimates on the deflections by the EGMF}
\label{app:egal}

Since the EGMF is not included in the simulations of the main text, we estimate the expected angular deflections induced by the Extragalactic Magnetic Field (EGMF). Unlike the GMF, the EGMF is poorly constrained, with strength estimates ranging from $10^{-15}$\,G in voids to $\mu$G in cluster centers. We adopt a turbulent magnetic field model characteristic of the Local Supercluster structures, assuming a conservative RMS field strength of $B_{\rm rms} = 1$\,nG and a coherence length of $L_{\rm coh} \approx 0.2$\,Mpc. 

Given that the source distance $D$ is typically much larger than $L_{\rm coh}$, the propagation is treated as a diffusive process. In the small-angle limit, the 1D projected scale parameter $\sigma$ for the magnetic scattering of a particle with charge $Z$ and energy $E$ is derived from \cite{Harari:2002dy} as:
\begin{equation}
    \sigma \simeq 0.8^\circ \, Z \, \left(\frac{100\,\text{EeV}}{E}\right) \, \left(\frac{B_{\rm rms}}{1\,\text{nG}}\right) \, \sqrt{\frac{D}{10\,\text{Mpc}}} \, \sqrt{\frac{L_{\rm coh}}{1\,\text{Mpc}}}
    \label{eq:egmf_deflection}
\end{equation}
We note that Equation \ref{eq:egmf_deflection} serves as a lower bound; for highly deflected trajectories, the small-angle approximation breaks down and true deflections may be larger. The total 2D angular separation $\theta$ between the true source and the observed arrival direction follows a Rayleigh distribution governed directly by this scale parameter $\sigma$.

A key uncertainty in UHECR propagation is the disintegration history of the parent nucleus. If fragmentation occurs significantly after emission, the kinematic and magnetic separation is reduced compared to source-fragmentation scenarios. However, because photo-disintegration approximately preserves the Lorentz factor, the rigidity $R = E/Z$ of a secondary fragment remains coupled to the mass-to-charge ratio of the primary. While exact rigidity is not conserved at every intermediate step of the cascade, the $A/Z \approx 2$ ratio of stable endpoints ensures that the total deflection scale remains bounded by the primary’s magnetic trajectory.

\begin{figure}[htbp]
    \centering
    \includegraphics[width=\textwidth]{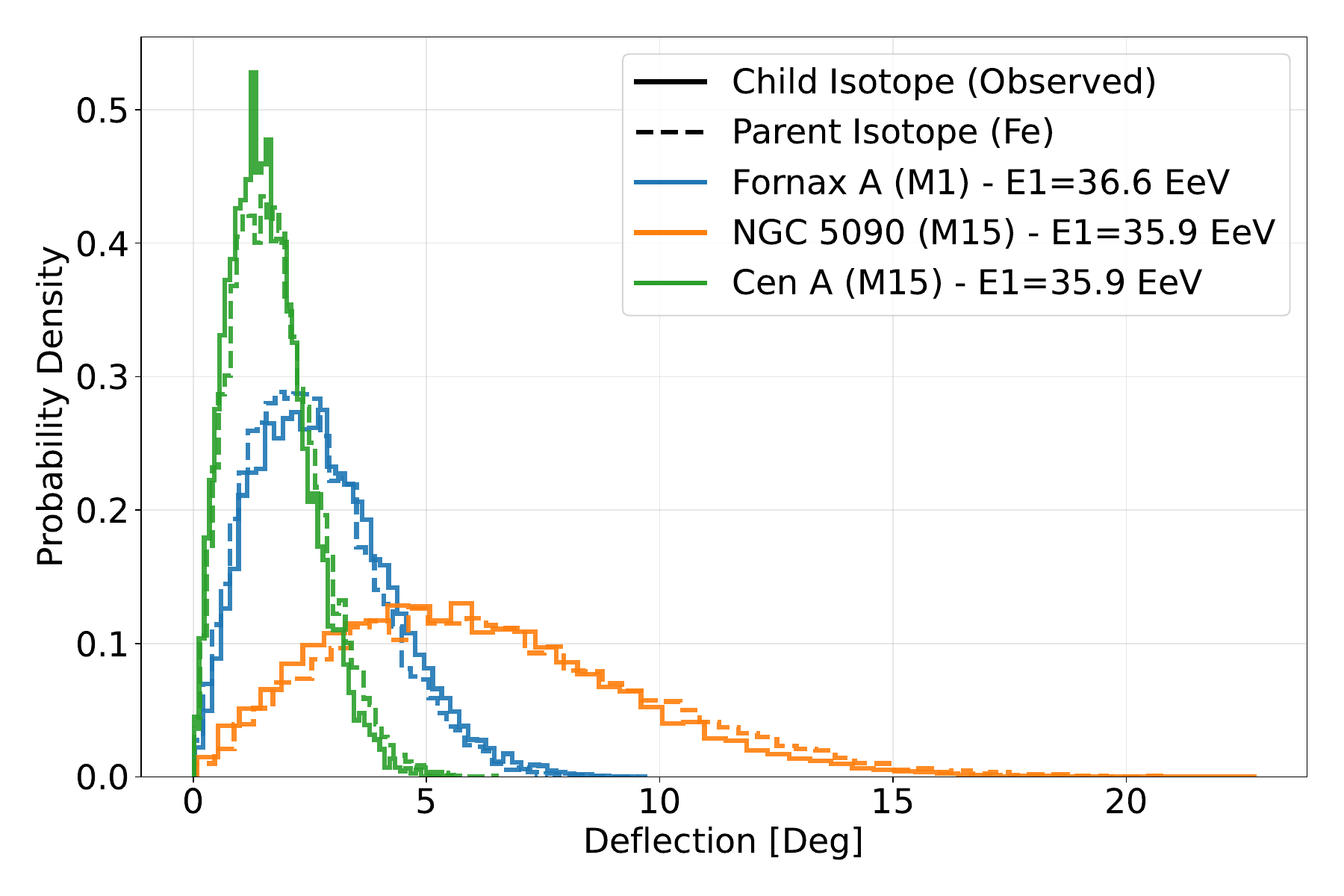}
    \caption{Analytical EGMF deflections ($B_{\rm rms}=1$\,nG, $L_{\rm coh}=0.2$\,Mpc) for the lower-energy ($E_1$) event of multiplets from three representative sources. Plotting only $E_1$ bounds the maximum expected deflection. We compare propagation as the observed ``Child'' nucleus (solid) versus an Iron ``Parent'' (dashed) injected with energy $E_{\rm Fe} = E_{\rm obs} \cdot (56/A_{\rm obs})$ to conserve the Lorentz factor.}
    \label{fig:egmf_distributions}
\end{figure}

Figure~\ref{fig:egmf_distributions} demonstrates that the backtracking results are robust against these disintegration histories for the ensemble of propagated particles. As shown in Table~\ref{tab:egmf_results}, the predicted EGMF deflections generally explain the residual offsets. For instance, the Helium doublets associated with Fornax~A (IDs 1 and 8) show a mean deflection of $\sim 2^\circ$, matching the distance ($\Psi_{\text{src}}$) to the source. For the NGC\,5090 -- Cen\,A ambiguity, the substantial EGMF deflection predicted, especially for NGC\,5090 ($\gtrsim 5^\circ$), makes it impossible to spatially distinguish between these two potential sources. The larger deflection required for the association with NGC 5090 ($\delta_{GMF} \approx 15^\circ$) compared to Centaurus A ($\delta_{GMF} \approx 4^\circ$) for ID 25 is a direct consequence of its significantly greater distance.

\begin{table}[htbp]
    \centering
    \caption{Estimated EGMF deflections for both events ($E_1$ and $E_2$) of the statistically associated independent doublets ($B_{\rm rms}=1$\,nG, $L_{\rm coh}=0.2$\,Mpc). $\Psi_{\text{src}}$ denotes the angular distance between the source and the backtracked doublet centroid. Multiplets are sorted by decreasing pre-trial significance.}
    \label{tab:egmf_results}
    \vspace{0.5cm} 
    \rotatebox{90}{%
        \resizebox{0.92\textheight}{!}{%
        \begin{tabular}{l l c c c c c c c c c c c c c c c c}
            \hline
            & & & & \multicolumn{7}{c}{\textbf{Child (Observed)}} & \multicolumn{7}{c}{\textbf{Potential parent (Iron)}} \\
            \cline{5-11} \cline{12-18}
            & & & & & \multicolumn{3}{c}{\textbf{Event 1 ($E_1$)}} & \multicolumn{3}{c}{\textbf{Event 2 ($E_2$)}} & & \multicolumn{3}{c}{\textbf{Event 1 ($E_1$)}} & \multicolumn{3}{c}{\textbf{Event 2 ($E_2$)}} \\
            \cline{6-8} \cline{9-11} \cline{13-15} \cline{16-18}
            \textbf{ID} & \textbf{Source} & \textbf{Dist [Mpc]} & \textbf{$\Psi_{\text{src}}$ [$^\circ$]} & \textbf{Iso} & \textbf{E [EeV]} & \textbf{$\delta_{\text{mean}}$ [$^\circ$]} & \textbf{$\sigma_{\delta}$ [$^\circ$]} & \textbf{E [EeV]} & \textbf{$\delta_{\text{mean}}$ [$^\circ$]} & \textbf{$\sigma_{\delta}$ [$^\circ$]} & \textbf{Iso} & \textbf{E [EeV]} & \textbf{$\delta_{\text{mean}}$ [$^\circ$]} & \textbf{$\sigma_{\delta}$ [$^\circ$]} & \textbf{E [EeV]} & \textbf{$\delta_{\text{mean}}$ [$^\circ$]} & \textbf{$\sigma_{\delta}$ [$^\circ$]} \\
            \hline
            1 & Fornax A & 26.1 & 2.36 & He & 36.6 & 2.81 & 1.47 & 53.4 & 1.92 & 1.01 & Fe & 512.4 & 2.59 & 1.35 & 747.6 & 1.79 & 0.93 \\
            8 & Fornax A & 26.1 & 2.39 & He & 33.0 & 3.10 & 1.63 & 42.3 & 2.41 & 1.26 & Fe & 462.0 & 2.89 & 1.51 & 592.2 & 2.25 & 1.17 \\
            7 & Fornax A & 26.1 & 2.91 & Be & 43.4 & 4.71 & 2.49 & 45.5 & 4.52 & 2.35 & Fe & 270.0 & 4.94 & 2.56 & 283.1 & 4.69 & 2.43 \\
            9 & CGCG 114-025 & 75.8 & 19.67 & He & 33.1 & 5.26 & 2.73 & 87.5 & 1.99 & 1.03 & Fe & 463.4 & 4.91 & 2.59 & 1225.0 & 1.87 & 0.97 \\
            15 & NGC 5090 & 51.0 & 3.97 & Li & 35.9 & 6.00 & 3.14 & 38.9 & 5.54 & 2.89 & Fe & 287.2 & 6.48 & 3.39 & 311.2 & 5.94 & 3.08 \\
            24 & NGC 5090 & 51.0 & 9.18 & O & 39.9 & 14.36 & 7.56 & 67.4 & 8.49 & 4.48 & Fe & 139.7 & 13.33 & 6.98 & 235.9 & 7.80 & 4.14 \\
            4 & Fornax A & 26.1 & 11.35 & N & 40.6 & 8.80 & 4.61 & 41.4 & 8.66 & 4.54 & Fe & 162.4 & 8.12 & 4.24 & 165.6 & 8.07 & 4.22 \\
            27 & Fornax A & 26.1 & 24.01 & N & 33.1 & 10.89 & 5.62 & 47.2 & 7.58 & 3.95 & Fe & 132.4 & 10.15 & 5.30 & 188.8 & 7.01 & 3.67 \\
            17 & Fornax A & 26.1 & 25.49 & O & 36.8 & 11.16 & 5.82 & 39.8 & 10.34 & 5.42 & Fe & 128.8 & 10.36 & 5.38 & 139.3 & 9.51 & 4.96 \\
            25 & NGC 0315 & 74.0 & 14.63 & N & 33.0 & 18.31 & 9.56 & 38.5 & 15.60 & 8.18 & Fe & 132.0 & 17.11 & 8.88 & 154.0 & 14.49 & 7.49 \\
            5 & M87 & 19.4 & 16.00 & Be & 34.7 & 5.10 & 2.66 & 62.5 & 2.82 & 1.47 & Fe & 215.9 & 5.32 & 2.77 & 388.9 & 2.96 & 1.54 \\
            0 & NGC 0315 & 74.0 & 23.86 & N & 32.9 & 18.45 & 9.58 & 38.0 & 15.92 & 8.32 & Fe & 131.6 & 16.99 & 8.79 & 152.0 & 14.77 & 7.71 \\
            18 & NGC 5090 & 51.0 & 26.22 & O & 61.6 & 9.29 & 4.84 & 88.9 & 6.46 & 3.35 & Fe & 215.6 & 8.71 & 4.53 & 311.2 & 5.99 & 3.14 \\
            23 & Fornax A & 26.1 & 31.55 & N & 32.6 & 11.02 & 5.79 & 33.0 & 10.83 & 5.68 & Fe & 130.4 & 10.32 & 5.35 & 132.0 & 10.14 & 5.27 \\
            16 & M87 & 19.4 & 22.25 & Fe & 45.9 & 25.08 & 13.04 & 52.5 & 21.89 & 11.48 & Fe & 45.9 & 25.05 & 12.96 & 52.5 & 21.91 & 11.62 \\
            12 & M87 & 19.4 & 26.40 & Fe & 35.8 & 31.99 & 16.64 & 39.1 & 29.25 & 15.24 & Fe & 35.8 & 32.05 & 16.82 & 39.1 & 29.22 & 15.33 \\
            3 & 2MASX J1254 & 69.1 & 34.17 & O & 33.0 & 20.26 & 10.59 & 41.2 & 16.13 & 8.51 & Fe & 115.5 & 18.81 & 9.75 & 144.2 & 15.11 & 7.90 \\
            10 & M87 & 19.4 & 25.77 & Fe & 39.5 & 28.84 & 15.05 & 56.7 & 20.27 & 10.49 & Fe & 39.5 & 29.02 & 15.12 & 56.7 & 20.21 & 10.54 \\
            6 & 2MASX J1254 & 69.1 & 37.24 & O & 33.7 & 19.79 & 10.29 & 38.5 & 17.19 & 8.98 & Fe & 118.0 & 18.38 & 9.46 & 134.8 & 16.08 & 8.36 \\
            11 & M87 & 19.4 & 31.03 & O & 33.7 & 10.59 & 5.46 & 33.8 & 10.49 & 5.51 & Fe & 118.0 & 9.79 & 5.15 & 118.3 & 9.63 & 5.05 \\
            20 & M87 & 19.4 & 40.92 & Fe & 35.3 & 32.61 & 17.18 & 57.9 & 19.83 & 10.50 & Fe & 35.3 & 32.52 & 16.99 & 57.9 & 19.88 & 10.47 \\
            21 & Fornax A & 26.1 & 17.28 & O & 72.8 & 5.64 & 2.97 & 80.3 & 5.14 & 2.68 & Fe & 254.8 & 5.26 & 2.74 & 281.1 & 4.71 & 2.48 \\
            \hline
        \end{tabular}%
        }
    }
\end{table}

\section{Time delay of the doublets}
\label{app:time_delay}

The temporal and angular analysis for the identified multiplets is presented in Table~\ref{tab:detailed_timing}. Simulations of UHECR propagation indicate that typical turbulent scattering and magnetic path lengthening result in temporal spreads of $\sigma_t \sim 500$--$5,000$ years, depending on the rigidity and source distance. In contrast, the observed doublets arrive within a window of days ($\Delta T_{\text{obs}} \lesssim 15$ days). 

We calculated the theoretical angular separation ($\theta_{\text{req}}$) expected if the observed time delay were caused solely by the geometric path difference between two magnetically deflected trajectories. Using the approximation $\Delta T \approx \frac{L \theta^2}{2c}$, where $L$ is the distance to the source, the required physical separation is:
\begin{equation}
    \theta_{\text{req}} \approx \sqrt{\frac{2 c \Delta T_{\text{obs}}}{L}}
\end{equation}

The "Required Angle" ($\theta_{\text{req}}$) derived from these 15-day delays is on the order of $\sim 0.001^\circ$, which is negligible compared to the observed angular separation ($\theta_{\text{obs}} \approx 1^\circ$--$3^\circ$). This discrepancy implies that the observed angular separation cannot be attributed to large-scale magnetic deflection differences of particles emitted simultaneously; such a scenario would inevitably introduce time delays on the order of millennia. 

\begin{table}[htbp]
    \centering
    \caption{Detailed temporal and angular analysis for the identified multiplets.}
    \label{tab:detailed_timing}
    \resizebox{\textwidth}{!}{%
    \begin{tabular}{l l c c c c c c c}
        \hline
        \textbf{ID} & \textbf{Source} & \textbf{Iso} & \textbf{$\theta_{\text{obs}}$ [$^\circ$]} & \textbf{$\Delta T_{\text{obs}}$ [days]} & \textbf{$\theta_{\text{req}}$ [$^\circ$]} & \textbf{$\Delta \bar{T}_{\text{sim}}$ [yr]} & \textbf{$\sigma_{t,1}$ [yr]} & \textbf{$\sigma_{t,2}$ [yr]} \\
        \hline
        1 & Fornax A & He & 1.57 & 5.8 & 0.0011 & 897 & 816 & 459 \\
        8 & Fornax A & He & 1.58 & 12.0 & 0.0016 & 1825 & 645 & 743 \\
        7 & Fornax A & Be & 2.49 & 6.0 & 0.0011 & 1412 & 1216 & 1141 \\
        9 & CGCG 114-025 & He & 1.80 & 6.3 & 0.0007 & 2486 & 2029 & 510 \\
        15 & NGC 5090 & Li & 1.67 & 8.7 & 0.0010 & 3316 & 2662 & 3026 \\
        27 & Fornax A & N & 1.98 & 11.6 & 0.0016 & 4201 & 2290 & 3080 \\
        24 & NGC 5090 & O & 2.60 & 11.0 & 0.0011 & 6216 & 7055 & 3721 \\
        4 & Fornax A & N & 1.81 & 6.9 & 0.0012 & 2548 & 2133 & 2350 \\
        17 & Fornax A & O & 1.39 & 13.7 & 0.0017 & 3832 & 3231 & 3597 \\
        25 & NGC 0315 & N & 2.85 & 9.9 & 0.0009 & 4538 & 3772 & 3240 \\
        5 & M87 & Be & 1.70 & 0.2 & 0.0002 & 4532 & 1465 & 1200 \\
        23 & Fornax A & N & 2.34 & 14.5 & 0.0018 & 2730 & 2392 & 2446 \\
        0 & NGC 0315 & N & 1.10 & 1.0 & 0.0003 & 4628 & 3766 & 4229 \\
        18 & NGC 5090 & O & 2.39 & 9.2 & 0.0010 & 3582 & 2366 & 3562 \\
        26 & UGC 7360 & Fe & 2.92 & 1.3 & 0.0005 & 20598 & 5206 & 17420 \\
        16 & M87 & Fe & 2.90 & 5.8 & 0.0013 & 6577 & 6911 & 4318 \\
        3 & 2MASX J1254 & O & 2.34 & 6.5 & 0.0007 & 10491 & 12943 & 5975 \\
        12 & M87 & Fe & 2.97 & 7.2 & 0.0015 & 7417 & 7849 & 6482 \\
        6 & 2MASX J1254 & O & 1.59 & 0.8 & 0.0003 & 4178 & 3440 & 3911 \\
        10 & M87 & Fe & 2.55 & 3.8 & 0.0011 & 8872 & 8100 & 8203 \\
        20 & M87 & Fe & 2.92 & 4.1 & 0.0011 & 3534 & 4099 & 2416 \\
        14 & M87 & Fe & 2.77 & 3.0 & 0.0009 & 9739 & 8889 & 9230 \\
        11 & M87 & O & 2.71 & 12.0 & 0.0019 & 1748 & 1523 & 1556 \\
        21 & Fornax A & O & 2.89 & 1.8 & 0.0006 & 1392 & 1072 & 1319 \\
        \hline
    \end{tabular}
    }
    \vspace{1ex}
    \parbox{\textwidth}{\footnotesize \textbf{Note:} Sorted by their association significance. 
    \textbf{$\theta_{\text{obs}}$}: Observed angular separation between doublet events [$^\circ$].
    \textbf{$\Delta T_{\text{obs}}$}: Observed time delay [days].
    \textbf{$\theta_{\text{req}}$}: Angle required to produce $\Delta T_{\text{obs}}$ via geometric delay [$^\circ$].
    \textbf{$\Delta \bar{T}_{\text{sim}}$}: Simulated mean arrival time difference [yr].
    \textbf{$\sigma_{t,1}$ / $\sigma_{t,2}$}: Simulated temporal dispersion [yr].}
\end{table}

\acknowledgments

I would like to thank Dmitriy Kostunin and Emma Kun for their valuable contributions during the early discussions regarding the analysis of the Pierre Auger Observatory data. I am also grateful to Vitor de Souza and Rita C. Anjos for their insightful discussions and continued support throughout the development of this work. I also thank Ralph Engel for the discussions during the 60th Rencontres de Moriond VHEPU in Italy this year, and the conference organization for supporting the trip. I would like to thank the State of Goiás Research Support Foundation (Fundação de Amparo à Pesquisa do Estado de Goiás - FAPEG) for financial support on this same trip. I acknowledge the use of Google Gemini 3 Pro as a support tool for language editing and assistance in optimizing specific segments of the analysis code. All AI-generated suggestions were independently verified and validated by the author.

% Bibliography

%% [A] Recommended: using JHEP.bst file
%% \bibliographystyle{JHEP}
%% \bibliography{biblio.bib}

%% or
%% [B] Manual formatting (see below)
%% (i) We suggest to always provide author, title and journal data or doi:
%% in short all the informations that clearly identify a document.
%% (ii) please avoid comments such as "For a review'', "For some examples",
%% "and references therein" or move them in the text. In general, please leave only references in the bibliography and move all
%% accessory text in footnotes.
%% (iii) Also, please have only one work for each \bibitem.

\bibliographystyle{JHEP}
\bibliography{biblio}

\end{document}